\documentclass{svmult}

\usepackage{bm}
\usepackage{mathptmx}       % selects Times Roman as basic font
\usepackage{helvet}         % selects Helvetica as sans-serif font
\usepackage{courier}        % selects Courier as typewriter font
\usepackage{type1cm}        % activate if the above 3 fonts are
\usepackage{makeidx}         % allows index generation
\usepackage{graphicx}        % standard LaTeX graphics tool
\usepackage{multicol}        % used for the two-column index
\usepackage[bottom]{footmisc}% places footnotes at page bottom
\newcommand{\bea}{\begin{eqnarray}}
\newcommand{\eea}{\end{eqnarray}}
\makeindex             % used for the subject index
\begin{document}

\title*{Photon Production in Hot and Dense Strongly Interacting Matter}
\author{Charles Gale}
\institute{Department of Physics, McGill University\\ 3600 rue University, Montreal, QC, Canada H3A 2T8} %\email{gale@physics.mcgill.ca}}
\maketitle 
\abstract{This text is meant as an introduction to the theoretical physics of photon emission in hot and dense strongly interacting matter, the principal application being relativistic nuclear collisions. We shall cover some of the results and techniques appropriate for studies at SPS, RHIC, and LHC energies.}
\section{Introduction}
The formation, observation, and study of the quark-gluon plasma (QGP) is one of the important goals of contemporary nuclear physics. Put more broadly, this program pursues an ambitious agenda which consists of mapping out the properties of a relativistic many-body system possibly out of equilibrium. Another way to state this goal is perhaps more intuitive: One seeks to explore the phase diagram of QCD (QuantumChromoDynamics: The theory of the strong interaction). This motivation has driven an energetic experimental program for several decades, at facilities which have included the AGS (Alternating Gradient Synchrotron, at Brookhaven National Laboratory), and the SPS (Super Proton Synchrotron, at CERN in Geneva, Switzerland). This research is in full bloom today, with experiments being carried on at RHIC (the Relativistic Heavy Ion Collider, at Brookhaven National Laboratory), and soon to be started at the LHC (the Large Hadron Collider, at CERN). In parallel, lower energy experiments continue at the GSI (Helmholtzzentrum f\"ur Schwerionenforschung, in Darmstadt, Germany) and will also be part of the FAIR facility at the same site. 

Many variables have been proposed as probes of the QGP \cite{vol1}, but we will concentrate here on electromagnetic observables, more specifically on real photons.  This text will concentrate on the theoretical techniques to calculate electromagnetic emissivities: The main experimental results and procedures are discussed elsewhere \cite{itzhak}. 

Three decades of relativistic heavy ion physics have made abundantly clear that progress in this field is usually associated with the measurement and interpretation of more than one experimental observable: The many-body nature of this problem is manifest. Moreover, complementary observables are not restricted to those of a hadronic nature. In fact, as they interact only electromagnetically, real and virtual photons have the potential to probe the entire dynamical history of heavy ion collisions. The mean free path of photons being much larger than typical nuclear scales, they will leave the interaction zone unscathed and therefore constitute penetrating probes of the systems under scrutiny. However, the absence of final state interactions is both an advantage and an inconvenient, the latter being small production rates ($\alpha \ll \alpha_{s}$). Furthermore, photons will be emitted throughout the entire history of the nuclear collisions: A quantitative interpretation of the asymptotic spectrum  requires an equivalent understanding of the signal, together with a thorough understanding of the background.

This text is arranged as follows: A brief outline of the formalism to calculate electromagnetic emissivities is given, followed by a discussion of photon production at the SPS, then at RHIC. At this latter energy scale, we review the electromagnetic signature of jets, together with the information they reveal on the strongly interacting medium. It is important to insist here that this text should be considered an introduction to some of the principles germane to the theoretical evaluation of electromagnetic emission rates and yields; it is not a detailed review.

\section{Photon production: General principles}

\label{sec:1}
A formalism that lends itself well to the calculation of thermal radiation of electromagnetic quanta is that of finite-temperature field theory. Quite generally, if one considers two hadronic states $| i \rangle$ and $| f \rangle$ that differ only by the absorption or the emission of a photon, the transition rate between those two may be written as \cite{KGbook} 
\bea
R_{f i} = \frac{| S_{fi}|^{2}}{t V}
\eea
where $t V$ is the proper four-volume, the $S$-matrix element is $S_{f i} = \langle f | \int d^{4}x\, \hat{J}_{\mu} (x) A^{\mu} (x) |i \rangle$, $\hat{J}_{\mu}$ is the hadronic electromagnetic current operator, and $A_{\mu} (x)$ is the photon field. After writing a free vector field as $A^{\mu} (x) = \epsilon^{\mu} \left( \E^{i k \cdot x} + \E^{- i k \cdot x} \right)/\sqrt{2 \omega V}$ ($\epsilon^{\mu}$ is a polarization vector and $k = (\omega, {\bf k})$), and noting that the matrix element is translation-invariant, one arrives at
\bea
R_{f i} = - \frac{g^{\mu \nu}}{2 \omega V} (2 \pi)^{4}\left[ \delta (p_{i} + k - p_{f}) + \delta (p_{i} - k - p_{f} ) \right] \langle f | \hat{J}_{\mu} (0) | i \rangle \langle i | \hat{J}_{\nu} (0) | f \rangle
\eea
where one delta function is relevant for the absorption process, and the other for emission. Taking a thermal average over initial states and summing over final states, one may write the emission rate in terms of a spectral function:
\bea
\omega \frac{d^{3} R}{d^{3} k} &=&  \frac{g^{\mu \nu}}{(2 \pi)^{3}} \pi f_{\mu \nu}^{-} (\omega, {\bf k})\hfill  \nonumber\\
f_{\mu \nu}^{-} (\omega, {\bf k})& = &- \frac{1}{Z} \sum_{i, f} \E^{- \beta {K}_{i}} (2 \pi)^{3} \delta (p_{i} - p_{f} - k) \langle f| \hat{J}_{\mu} (0) | i \rangle \langle i | \hat{J}_{\nu} (0) | f \rangle
\eea
where $Z = \sum_{i} \E^{- \beta {K}_{i}}$, with ${K} = {H} - \mu {N}$. 
Going from the current-current correlation function to the field-field correlator with the equation of motion $\partial^{\mu} \partial_{\mu} A_{\nu} (x) = J_{\nu} (x)$, and using the Schwinger-Dyson equation, one can write the emission rate as
\bea
\omega \frac{d^{3} R}{d^{3} k} = - \frac{g^{\mu \nu}}{(2 \pi)^{3}} \frac{1}{\E^{\beta \omega} - 1}{\rm Im} \Pi^{' {\rm R}}_{\mu \nu} (\omega, {\bf k})
\eea 
where $\Pi^{' {\rm R}}_{\mu \nu}$ is the finite-temperature retarded improper self-energy \cite{KGbook,FW} which, to leading order in the electromagnetic fine structure constant $\alpha$, is equal to the finite-temperature retarded proper photon self-energy. Therefore
\bea
\omega \frac{d^{3} R}{d^{3} k} = - \frac{g^{\mu \nu}}{(2 \pi)^{3}} \frac{1}{\E^{\beta \omega} - 1}{\rm Im} \Pi^{ {\rm R}}_{\mu \nu} (\omega, {\bf k}) + {\cal O} (\alpha^{2})
\label{impart}
\eea
From Eq. (\ref{impart}) it is immediately clear why photon measurements in relativistic heavy ion collisions have the potential to reveal precious information about the strong interacting system. This equation is perturbative in the electromagnetic sector but is {\bf exact} to all orders in the strong coupling. Photon measurements (and in fact electromagnetic observables in general) have the potential to carry direct information about the in-medium photon self-energy at finite temperatures and densities: A quantity which is currently not calculable.  Bear in mind however that this information needs to be convolved with the time evolution of the colliding system.

\subsection{Photons from thermal hadrons}
\label{thermalhadrons}

In a heavy ion collision at relativistic energies, a plethora of mesons are produced, from the light to the intermediate mass scale \cite{PBM}. To model their interaction, a Massive Yang-Mills approach can be used \cite{MYM} to write down a Lagrangian:  
\bea
{\cal L} &=& \textstyle{\frac{1}{8}} F_\pi^2 {\rm Tr} D_\mu U D^\mu U^\dag + \frac{1}{8}
F_\pi^2 {\rm Tr} M (U + U^\dag -2)\nonumber \\& &  - \textstyle{\frac{1}{2}}
{\rm Tr} \left(F_{\mu \nu}^L
{F^L}^{\mu \nu} + F_{\mu \nu}^R {F^R}^{\mu \nu} \right) + m_0^2 {\rm Tr}
\left(A_\mu^L {A^L}^{\mu } + A_\mu^R {A^R}^\mu\right)+
\gamma {\rm Tr}F_{\mu \nu}^L U F^{R \mu \nu}U^\dag\nonumber 
\\& & -i\xi {\rm Tr}\left(D_\mu UD_\nu U^\dag F^{L \mu \nu}
+D_\mu U^\dag D_\nu U F^{R \mu \nu}\right)
\label{Lmym}
\eea
The fields are
\bea
\lefteqn{U = \exp \left( \frac{2 i}{F_\pi} \sum_i \frac{\phi_i
\lambda_i}{\sqrt{2}}\right) = \exp\left( \frac{2 i}{F_\pi} 
\mbox{\boldmath $\phi$} \right)}\ \nonumber \\
 & & A_\mu^L = \textstyle{\frac{1}{2}}(V_\mu + A_\mu)\  \nonumber \\
 & & A_\mu^R = \textstyle{\frac{1}{2}}(V_\mu - A_\mu)\  \nonumber \\
 & & F_{\mu \nu}^{L, R}  = \partial_\mu A_\nu^{L, R} - \partial_\nu A_{\mu}^{L, R} -
i g_0 \left[A_{\mu}^{L, R}, A_\nu^{L, R} \right]\ \nonumber \\
 & & D_\mu U = \partial_\mu U - i g_0 A_\mu^L U + i g_0 U A_\mu^R\nonumber \\
 & & M = \frac{2}{3} \left[ m_K^2 + \frac{1}{2} m_\pi^2\right] - \frac{2}{\sqrt{3}}
(m_K^2 - m_\pi^2) \lambda_8\ 
\eea
Note also that $F_\pi$ = 135 MeV, and that $\lambda_i$ is a Gell-Mann matrix.  
$\phi, V_\mu$ and $A_\mu$ are, respectively, the pseudo-scalar, vector 
and axial vector meson field matrices, respectively (not the $A_{\mu}$ of the previous section). The unspecified parameters of the Lagrangian can be adjusted to reproduce adequate hadronic phenomenology. For example, the masses and widths of the $\rho$ and of the $a_1$ may be used as input \cite{song}. Importantly, the ratio of $D-$ to $S-$ wave in the final state of the $a_1 \to \rho \pi$ decay is determined experimentally to be $-0.062 \pm 0.020$ \cite{PDG}. Our choice of parameters yields a value of $-0.099$ \cite{TRG}. 

A practical and intuitive alternative to the field-theoretic approach to the calculation of photon emission rates is provided by relativistic kinetic theory. For example, the differential emission rate, $R$,  in the emission channel $1+ 2 \to 3 + \gamma$ can be obtained either from the imaginary part of the two-loop retarded photon self-energy at finite temperature, or equivalently from
\begin{eqnarray}
\omega \frac{d^{3} R}{d^{3}k} = 
\int \prod_{i=1}^{3}\frac{d^{3} p_{i}}{(2 \pi)^{3} 2 E_{i}} | {\cal M}|^{2}  \frac{f(E_{1}) f(E_{2}) \left(1 \pm f(E_{3})\right)}{2 (2 \pi)^{3}} (2 \pi)^{4} \delta^{4} (p_{1} + p_{2} - p_{3} -k) \label{rate} \nonumber \\
\end{eqnarray}
where ${\cal M}$ is the invariant scattering matrix element derived from the Lagrangian introduced previously. The plus (minus) sign is associated with a boson (fermion) hadron final state. The distribution functions, $f(E)$, are also  chosen appropriately. 

A study of the interaction vertices will generate all possible tree-level diagrams for the generic reactions $X + Y \to Z + \gamma$, $\rho \to Y + Z + \gamma$, and $K^* \to Y + Z + \gamma$. Here, the sets $\{ X, Y, Z \}$ represent all combinations of $\pi, \rho, K, K^*$ mesons which respect conservation of charge, isospin, strangeness, and $G$ parity defined for nonstrange mesons.   Using Eq. (\ref{rate}), the thermal production rates are obtained by coherently summing matrix elements in each channel; convenient parametrizations of the results are given in the appendix of Ref. \cite{TRG}. Note finally that the use of effective Lagrangians still necessitates the introduction of hadronic form factors at moderate and high momentum transfers, as hadrons are composite entities.   One expects a reduction in the 2-3 GeV region (in photon energy) that amounts to an approximate factor of 4 \cite{TRG,KLS}. A useful and effective form of the hadronic form factor is a standard dipole term: $F(t) = ( 2 \Lambda^{2}/(2 \Lambda^{2} - t))^{2}$, where $t$ is the usual Mandelstam variable. For $\Lambda$, an appropriate value for phenomenology is that of a  typical hadronic scale, $\Lambda$ = 1 GeV \cite{RappGale}.

In addition to the contribution from a thermal ensemble of interacting mesons, such as the one described above, there also will be photon radiation from channels involving baryons. A starting point here may be the hadronic many-body calculations used for the interpretation of low-mass dilepton measurements \cite{RW}. A key theoretical ingredient there is the in-medium spectral densities for the low-lying vector mesons, as can be understood from Eq. (\ref{impart}). The connection between the electromagnetic signal and the in-medium spectral density is made explicit through the famous Vector Dominance Model (VDM) \cite{VDM}: ${\rm Im} \Pi_{\mu \nu} = \sum_{i = \rho, \omega, \phi} (m_{i}^{4} /g_{i}^{2}) \, {\rm Im} D_{\mu \nu} (i)$. In effect, VDM identifies the hadronic electromagnetic current operator with the vector meson field operator. This approach has had considerable phenomenological success, and further refinements have also been considered \cite{VDM2}.

Another thermal source of radiation is introduced in the next section. After this is done, all purely thermal electromagnetic emissivities can be compared at a same temperature. 

\subsection{Photons from thermal partons}

\begin{figure}[htb]
\begin{center}
\includegraphics[width=10.0cm]{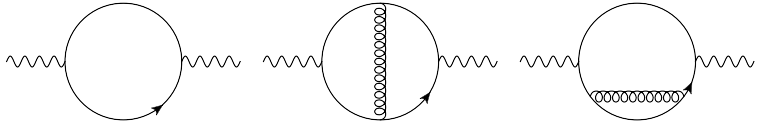}
\end{center}
\caption{Contributions to the photon self-energy in QCD, at one and two loops.}
\label{one-two-loops}
\end{figure}

\begin{figure}[htb]
\begin{center}
\includegraphics[width=3.5cm]{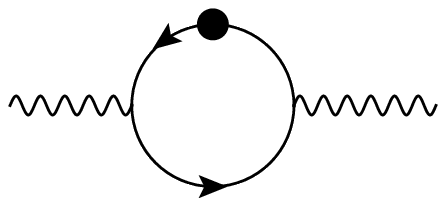}\hspace*{2cm}
\parbox{3cm}{\vspace*{-1.4cm}\includegraphics[width=3cm]{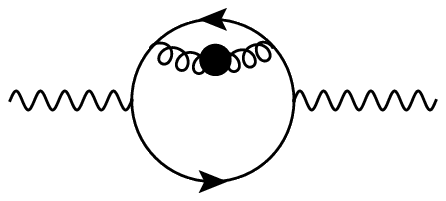}\\
\includegraphics[width=3cm]{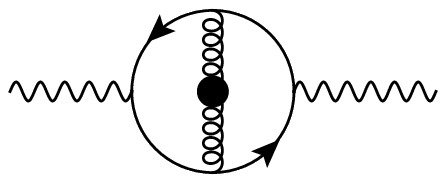}}
\end{center}
\caption{Contributions to the retarded, finite-temperature self-energy of the photon at one loop (left panel), and at two loops (right panel). The dark circle on the quark and gluon propagators represent the use of effective propagators, resumed using the Hard Thermal Loops effective theory. }
\label{oneloop}
\end{figure}

The calculation of the thermal emissivity of an ensemble of thermal partons proceeds similarly to cases discussed previously. The key ingredient is the retarded, in-medium, photon self-energy evaluated at finite-temperature. In this context, it is instructive to follow the chronological order of the different theoretical investigations. Early calculations first proceeded with an expansion in order of the loop topologies. It is revealing to spend some time on the rate for photon emission obtained from one- and two-loop self-energies \cite{KLS}. This discussion follows that in Ref. \cite{KLS}. Taking the imaginary part of the self-energies shown in Figure \ref{one-two-loops} will produce the annihilation and Compton channels for photon production which involve gluons, quarks, and antiquarks. Note that if the photon is virtual those cuts will also yield processes like $q \bar{q} \to \gamma^{*}$ and ${\cal O} (g^{2})$ corrections to it \cite{majgale}, all of which disappear when the photon goes on-shell. 
Taking the imaginary part of the QCD diagrams in Figure \ref{one-two-loops} yields the rate of Eq. (\ref{rate}) which, taking the Boltzmann limit for the distribution functions, may be integrated to yield
\bea
\omega {\frac{d^{3} R}{d^{3}k}}^{\rm annihilation} &=& \frac{5}{9} \frac{\alpha \alpha_{s}}{3 \pi^{2}} T^{2} \E^{-\beta \omega} \left[ \ln\left(\frac{4 \omega T}{k_{c}^{2}}\right) + \delta_{1}\right]\nonumber \\
\omega {\frac{d^{3} R}{d^{3}k}}^{\rm Compton} &=& \frac{5}{9} \frac{\alpha \alpha_{s}}{6 \pi^{2}} T^{2} \E^{-\beta \omega} \left[ \ln\left(\frac{4 \omega T}{k_{c}^{2}}\right) + \delta_{2}\right]
\label{rate-uv}
\eea
We have considered u and d quarks, and an implicit assumption here is that $\omega = E_{1} + E_{2} \gg T$, where $\omega$ is the photon energy. The values of the two constants $\delta_{1}$ and $\delta_{2}$ are known \cite{KLS}, and depend on whether quantum statistics are used or their classical limit. Importantly, $k_{c}$ is an infrared cutoff which regulates the rates: As the exchanged quark goes on shell, the well-known logarithmic divergence manifests itself if $k_{c} \to 0$. As is known from Hard Thermal Loops (HTL) analyses  a propagator must be dressed if the momentum flowing through is soft on a scale set by the temperature $T$ \cite{KGbook}. Consequently the infrared contribution to the imaginary part of the diagrams in Figure \ref{one-two-loops} will come from the diagram on the left-hand-side of Figure \ref{oneloop}. The blob on the quark line in the figure represents the HTL-resummed propagator, where thermal effects regularize the low momentum (infrared) behavior of the theory. 
For u and d quarks, this contribution may be written as 
\bea
\Pi^{\mu \nu} (p) = -6 \times \frac{5}{9} e^{2} T \sum_{k_{0}} \int \frac{d^{3}k}{(2 \pi)^{3}} {\rm Tr} \left[ {\cal G}^{*} (k) \gamma^{\mu} {\cal G} (p -k) \gamma^{\nu}\right]
\label{Pimunu}
\eea
where 
\bea
{\cal G}^{*} (k) = {\cal G}_{+}^{*} (k) \frac{\gamma_{0} - {\bf \hat{k}} \cdot \mbox{\boldmath $\gamma$}}{2} + {\cal G}_{-}^{*} (k) \frac{\gamma_{0} + {\bf \hat{k}}\cdot \mbox{\boldmath$ \gamma$}}{2}\ ,
\eea
the dressed propagator for a quark with momentum $k$, and
\bea
{\cal G}_{\pm}^*(k) = \left\{ -k_0 \pm |{\bf k}| + \frac{m_q^2}{|{\bf k}|}
\left[ Q_0(k_0/|{\bf k}|) \mp Q_1(k_0/|{\bf k}|) \right]\right\}^{-1}
\eea
The sum over $k_{0}$ in Eq. (\ref{Pimunu}) is a sum over Matsubara frequencies \cite{KGbook}, and the functions $Q_0$ and $Q_1$ are the Legendre functions of the second kind, 
namely
\bea
Q_0(z) &=& \frac{1}{2} \ln\left( \frac{1+z}{1-z} \right) \nonumber \\
Q_1(z) &=& z Q_0(z) -1
\eea
The effective quark mass is given below.  In the limit $g 
\rightarrow 0$  the bare quark propagator is recovered. Note also that $p_{0}$ will be analytically continued to $\omega$. 

In this language, the bare propagator is written as
\bea
{\cal G} (k) = g_{+} (k) \frac{\gamma_{0} - {\bf \hat{k}} \cdot \mbox{\boldmath $ \gamma$}}{2} + g_{-} (k) \frac{\gamma_{0} + {\bf \hat{k}} \cdot \mbox{\boldmath $\gamma$}}{2}
\eea
with
\bea
g_{\pm} (k) = \frac{1}{(-k_{0} \pm |{\bf k}|)}
\eea
As we know from HTL, the cutoff is soft on a scale set by $T$: We may choose it to lie in the interval between $m_{q}$ and $T$, where $m_{q}$ is the thermal quark mass: $m_{q}^{2} = g^{2}T^{2}/6$ (with $N_{c}$ = 3). Doing so, and adding the answer one gets to the ones from Eq. (\ref{rate-uv}), one arrives at \cite{KLS,BNNR}
\bea
\omega \frac{d^{3}R}{d^{3}k} = \frac{5}{9} \frac{\alpha \alpha_{s}}{2 \pi^{2}} T^{2}\E^{-\beta E} \ln \left(\frac{\delta_{3}}{g^{2}} \frac{\omega}{T}\right)
\label{KLSrate}
\eea
The value of the numerical constant $\delta_{3}$ is also known, and depends on whether Boltzmann or quantum statistics are invoked. It is remarkable that the final rate is cutoff-independent. 

Therefore, at this order in the loop expansion, a leading-order (in $\alpha_s$) contribution obtained from the imaginary part of the diagrams on the left panel of Figure \ref{oneloop} and of Figure \ref{one-two-loops} emerges . 
%The result of this calculation is \cite{KLS,BNNR}, for high energy photons:
%\bea
%{\rm Im} \Pi^{{\rm R} \mu}_{\mu} = 4 \pi \frac{5 \alpha \alpha_{s}}{9} T^{2} \left[ \ln\left( \frac{E_{\gamma} T}{m_{q}^{2}}\right) + \kappa \right]
%\eea
%where $\kappa$ is a known numerical constant, and $m_{q}^{2} = \pi \alpha_{s} C_{f} T^{2}$, with $C_{f} = (N_{c}^{2} - 1)/2 N_{c}$. For 2 flavors (u and d),  this equation is exact up to corrections of ${\cal O} (T/E)$. 
Notice the effect of using the HTL propagator: If the thermal mass is set to zero, the rate diverges. It is clear that HTL does its job of shielding the rate from the looming logarithmic singularity. 

\begin{figure}[htb]
\begin{center}
\includegraphics[width=7.0cm]{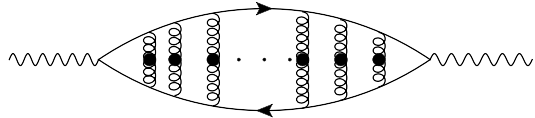}
\end{center}
\caption{A ladder diagram contribution to the retarded, finite-temperature,  in-medium, photon self-energy at order $\alpha_{s}$. The dark blobs represent effective HTL propagators.  Diagrams with crossed rungs, for example, have been shown to be sub-leading \cite{AMY}. }
\label{ladder}
\end{figure}

\begin{figure}[htb]
\begin{center}
\includegraphics[width=9.0cm]{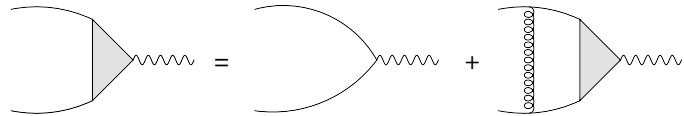}
\end{center}
\caption{A schematic resummation of ladder diagrams.}
\label{ladder2}
\end{figure}

Progress was made through subsequent evaluations involving higher order configurations in loop topology. In fact, taking cuts through the diagrams on the right panel of Figure \ref{oneloop} represented a somewhat involved calculation \cite{twoloops} that produced what first appeared to be a surprising result: For three colors and for 2 quark flavors, the two diagrams on the right-hand-side of Figure \ref{oneloop} contribute
\bea
{\rm Im}  \Pi^{{\rm R} \mu}_{\mu} = \frac{32}{3 \pi} \frac{5 \alpha \alpha_{s}}{9} \left[ \pi^{2} \frac{T^{3}}{\omega} +\omega T \right]
\eea
In other words, diagrams of a higher order in ``naive power  counting'' (${\cal O}(\alpha_s^2)$) (right panel of Figure \ref{oneloop}), contribute parametrically at order $\alpha_s$ as do diagrams of a less elaborate loop topology (left panel, same figure).  Technically, this is understood as follows: As the one-loop contribution produced a logarithmic singularity when the quark thermal  mass ($m_{q}$) tended to zero, the contribution from two-loop diagrams also become singular, but linearly. Specifically, the factor which blows up is $T^{2}/m_{q}^{2}$. That, multiplied by $\alpha_{s}^{2}$ (from the vertices), produces a net contribution which is ${\cal O} (\alpha_{s})$. Perhaps more physically, the photon formation time can be related to the virtuality of the quark emitting the photon  in a tree-level representation of the emission rate. Owing mainly to an uncertainty principle argument, quarks with low virtualities will take a long time to radiate photons. Long formation will then leave the process vulnerable to in-medium scattering, spoiling the eventual decoherence of the photon field. This is in effect a statement of the Landau-Pomeranchuk-Migdal effect (LPM) effect \cite{LPM}. Searching for all contributions at this order, a diagrammatic analysis revealed that, at leading order in $\alpha_{s}$, the only set of diagrams that contribute to the photon rate is obtained by summing over ladder diagrams \cite{AMY}, as that in Figure \ref{ladder}. 
The ladder contributions can be written as a linear integral equation and re-summed, as shown schematically in Figure \ref{ladder2}, to yield a finite result. Unfortunately, the net rate has not been amenable to a closed analytical form, but a convenient parametrization has been given in Ref. \cite{AMY2}.

\begin{figure}[htb]
\begin{center}
\includegraphics[angle=-90,width=6.0cm]{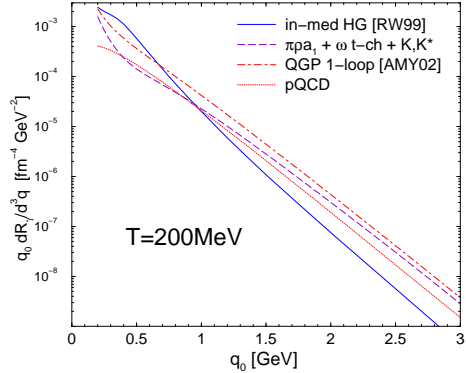}
\end{center}
\caption{A comparison of photon emission rates from different phases of the QCD phase diagram. The nature of the different sources is discussed in the text. From Ref. \cite{TRG}}
\label{photon_rates}
\end{figure}

Now, the purely thermal sources of photons may be compared against each other, in order to get some feeling for their relative importance. There will be radiation from a thermal ensembles of partons, and that from a thermal ensemble of hadrons from the confined sector of QCD. This comparison is done for a temperature of $T$ = 200 MeV, and shown in Figure \ref{photon_rates}. There, the rate for photon emission obtained from the hadronic many-body approach of Refs. \cite{RappGale,Rapp1,Rapp2} by considering the square of the virtual photon invariant mass and taking the limit $M^{2} \to 0$, is the solid line. The mesonic contributions discussed in Section \ref{thermalhadrons}, including hadronic form factors, are represented by the dashed curve. The photon rate from thermal partons \cite{AMY} obtained from taking the imaginary part of Figure \ref{ladder} is shown as the dashed-dotted line.  
At low momenta, the net spectrum is dominated by radiation from interacting hadrons. At a momentum scale of about 1 GeV, and above, the photons from thermal QCD shine on more brightly that those from competing sources. It is also interesting to note that the enhancement in rate provided by summing the infinite set of diagrams from Figure \ref{ladder} give a net boost of a rough factor of 2$\sim$2.5 over those obtained through Eq. (\ref{KLSrate}). At this stage, an optimistic summary is possible: If the energy/temperature is high enough, the photons from the plasma state of QCD will be a large component of the electromagnetic spectrum.        Also, the early conclusion that the hadronic and partonic phases have very similar electromagnetic emissivities \cite{KLS} might still be tenable: The QGP rates complete to leading order in $\alpha_{s}$ are within a factor of two of the sum of hadronic contributions (solid and dashed curves). Bear in mind, however, that as the temperature and density evolve in time, a net yield of photons will come from an integration through different landscapes of the QCD phase diagram. Also, there will be other sources of photons other than the ones discussed up to now, some of which will have a thermal component, some of which will not. The final spectrum has the potential to contain all of them. We shall explore these other channels in sections to follow. 

\section{Photons from nuclear collisions at CERN SPS energies}

\begin{figure}[htb]
\begin{center}
\includegraphics[angle=0,width=7.0cm]{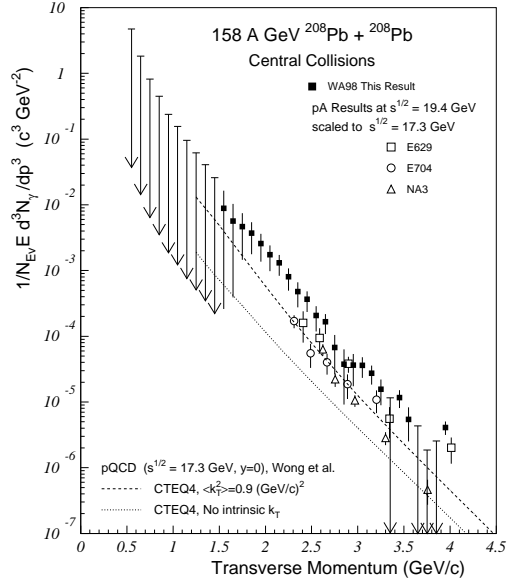}
\end{center}
\caption{The real photon spectrum spectrum as a function on transverse momentum, as measured by the WA98 collaboration. The pQCD estimates are from \cite{wongwang}.}
\label{wa98}
\end{figure}

The first measurements of photons at the CERN SPS were made by the WA80 \cite{WA80} and CERES \cite{ceres} collaborations. Unfortunately, only upper limits on an eventual thermal component could be set, owing in part to considerable systematic errors. Subsequently, the WA98 Collaboration identified a photon excess (over those produced by a superposition of nucleon-nucleon collisions) in central collisions of lead nuclei at a fixed-target energy of 158 AGeV \cite{WA98}. This data is shown in Figure \ref{wa98}. The collaboration compared its data to those obtained in different proton-induced reactions at a neighboring energy, scaled to that used by WA98. For transverse momenta such that $p_{T} \geq$ 1.5 GeV, this data shows a clear excess beyond what is expected from proton-nucleus collisions.
In the process of attempting to interpret theoretically the WA98 data, one must keep in mind that one photon source that is unavoidable is that associated with the primordial (during the very first instants of the nuclear collision) interactions of nucleons. As we shall see later, considerable sophistication has been reached in the calculation of this contribution at high energies. However, at $\sqrt{s} =$17.3 GeV, analyses are still subject to ambiguities. For example, the question of whether the partons colliding at these energies have some amount of intrinsic traverse momentum ($k_{T}$) has remained open: Attempts to extract some meaningful values from experiments have remained inconclusive \cite{patrick-photons}. Then, the absence of a consistent and successful theoretical framework at the level of proton-proton collisions has made the later assessment of nuclear contribution such as shadowing and {\it dynamical} $k_{T}$ broadening (the Cronin effect) challenging. In this context, the WA98 collaboration has recently shown \cite{wa98-II} an analysis of photons produced in proton-nucleus collisions at $\sqrt{s} =$ 17.4 GeV. However, this additional data has not  the level of precision necessary to remove theoretical uncertainties. It is a fair summary to state that the WA98 data has generated a flurry of theoretical activity \cite{wa98-theory}. We now concentrate on an approach which combines several of the elements discussed up to now \cite{TRG}.  

\begin{figure}[htb]
\begin{center}
\includegraphics[angle=-90,width=7.0cm]{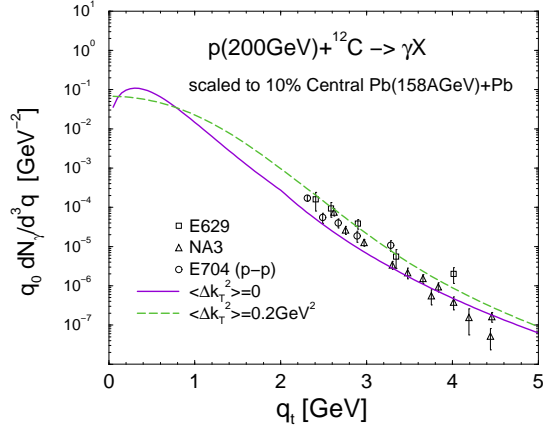}
\end{center}
\caption{Direct photon spectra from proton-nucleus collisions, scaled to central Pb + Pb collisions at SPS energies, with different values of the momentum broadening parameter $\langle \Delta k_{T}^{2}\rangle$, compared with experimental data \cite{pAdata}. From Ref. \cite{TRG}.}
\label{pA}
\end{figure}
The hadron and parton rates have been defined in earlier sections, and compared at a fixed temperature in Figure \ref{photon_rates}. As an accurate description of the direct photon in primordial nucleon-nucleon at SPS energies is still a matter unsettled, empirical scaling relations extracted from fits to experimental data \cite{dinesh-fits} are used. Starting from data acquired in pp collisions, one may naively extrapolate to a nucleus-nucleus (A + B) geometry with
\bea
\omega \frac{d^{3}N_{\gamma}}{d^{3}k}^{\rm prompt} &=& \omega \frac{d^{3}\sigma_{\gamma}}{d^{3}k}^{pp} A B T_{A B} (b)\nonumber\\
&=& \omega \frac{d^{3}\sigma_{\gamma}}{d^{3}k}^{pp} \frac{N_{\rm coll}}{\sigma_{pp}^{\rm in}}
\eea
where $T_{AB} (b)$ is the nuclear overlap function, $N_{\rm coll}$ is the number of nucleon-nucleon collisions, and $\sigma_{pp}^{\rm in}$ is the inelastic proton-proton collision cross section. The differential photon cross section is parametrized from experimental data \cite{TRG}. Since intrinsic $k_{T}$ smearing effects should in principle be contained in this parametrization, the dynamical smearing is generated by fitting an additional component to the available proton-nucleus data. This is tantamount to folding the parametrized spectrum with a Gaussian distribution:
\bea
f (k_{T}) = \frac{1}{\pi \langle \Delta k_{T}^{2}\rangle} \E^{-k_{T}^{2}/\langle \Delta k_{T}^{2}\rangle}  
\eea
One then obtains proton-nucleus photon spectra, which may be scaled \cite{WA98} to central Pb + Pb collisions at 158 AGeV; this is shown in Figure \ref{pA}. We may thus put all the ingredients together, and convolve our rates with a fireball evolution model which is also able to reproduce low and intermediate invariant mass dilepton measurement results \cite{rashu}. In this context, it is important to highlight the importance of the fireball's hadro-chemical composition in this analysis. At the SPS, a fit to several hadronic species has fleshed out a scenario where particles undergo chemical freeze-out before kinetic freeze-out \cite{PBM}. In the temperature zone between the two freeze-out values, the conservation of particle number demands the increase of chemical potentials: These will play an important role on photon production. In the hadronic phase, for example, the reaction $\pi \rho \to \pi \gamma$ is enhanced over the case with nil chemical potentials by a factor $\E^{3 \mu_{\pi}/T}$. It is therefore important to get a good quantitative understanding of both chemical potentials and transverse expansion: The latter blueshifts the spectrum towards harder momenta. 
\begin{figure}[htb]
\begin{center}
\includegraphics[angle=-90,width=6.5cm]{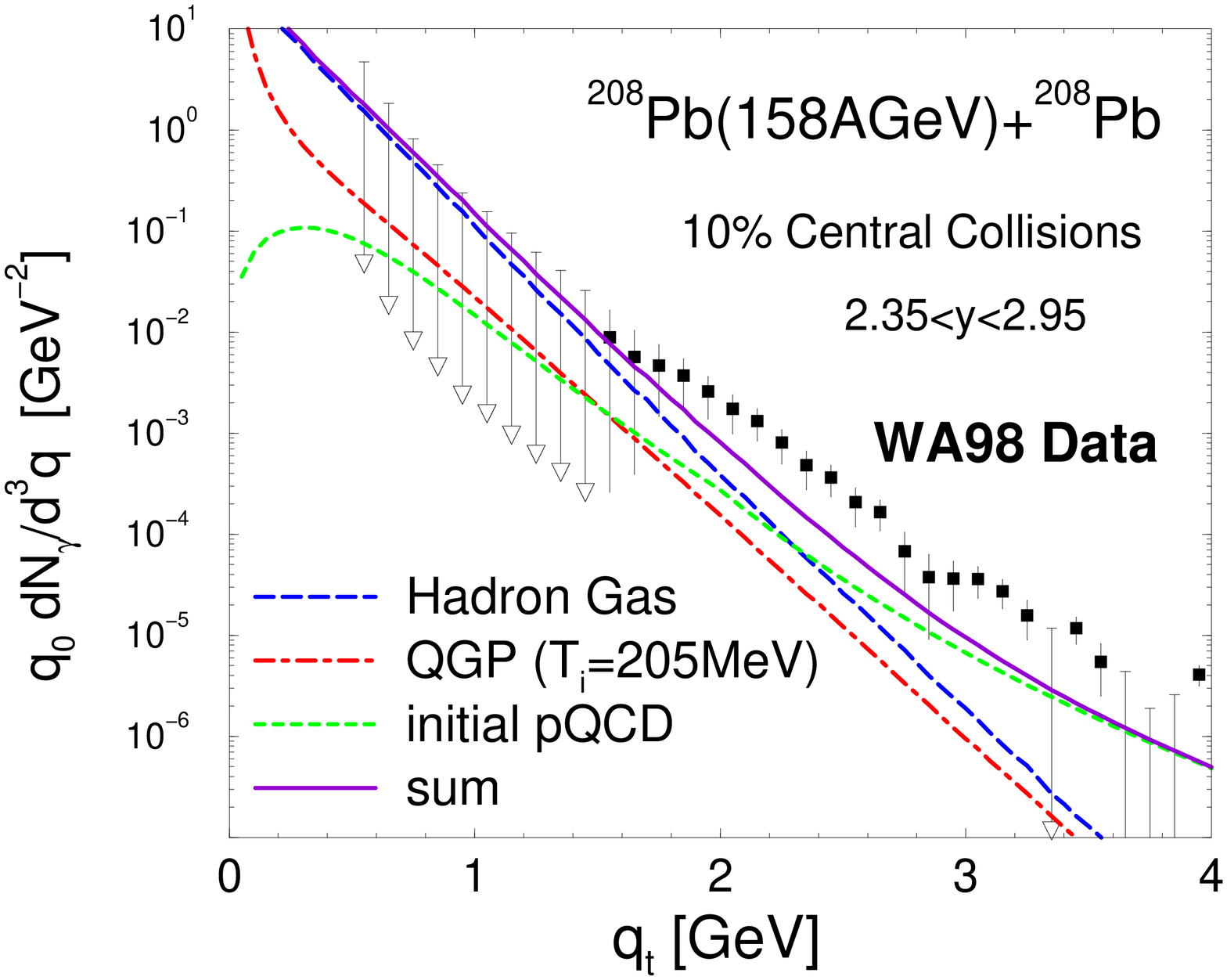}
\includegraphics[angle=-90,width=6.5cm]{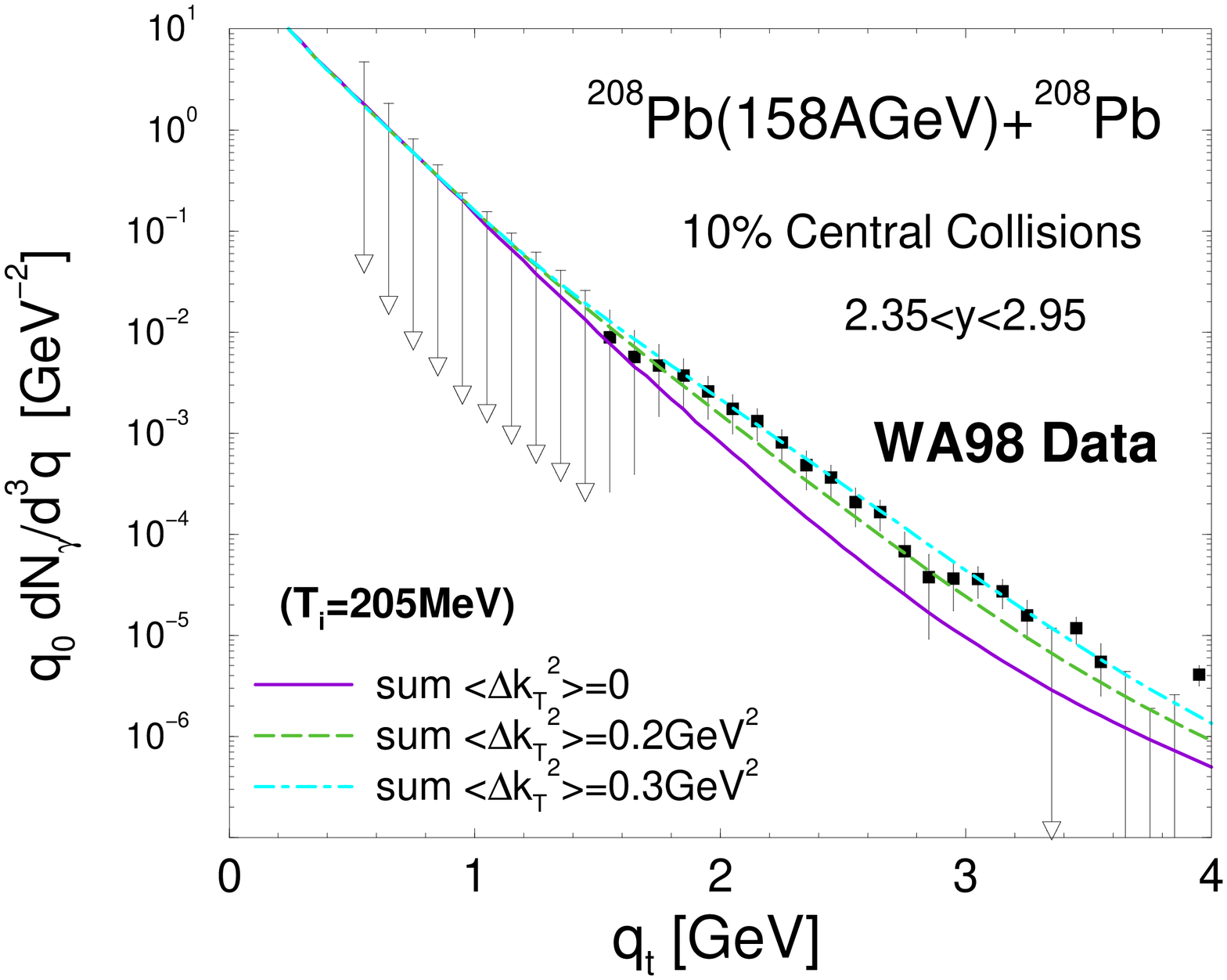}
\end{center}
\caption{Direct photon spectra from central Pb + Pb collisions at SPS energies. In the top panel, the contribution from different sources are shown separately: That from the hot ensemble of hadrons (long dashes), from the QCD plasma (dash-dot), and from primordial nucleon-nucleon collisions (short dashes). The sum is shown as a solid curve. In the bottom panel, the net signal is shown for different values of the momentum broadening parameter $\langle \Delta k_{T}^{2}\rangle$. The figure is from  Ref. \cite{TRG}.}
\label{trg-final}
\end{figure}
Bearing these empirical constraints in mind, one calculates the final spectrum shown in Figure \ref{trg-final}, together with the WA98 experimental data. 

The ingredients used here consist of a (schematic) fireball evolution which is nevertheless consistent with the measured hadrochemistry and with the hydrodynamic expansion global characteristics, of a comprehensive set of photon emissivity from a gas of hadrons, of partonic emissivities complete to leading order in $\alpha_{s}$, and of an estimate of the nuclear Cronin effect deduced from measurements made in proton-nucleus collisions generalized to nucleus-nucleus events. It is fair to write that one of the main conclusions emerging from this study is that the direct photons measured at CERN SPS energies are consistent with hadronic properties from he confined sector of QCD: The plasma contributes negligibly to the observed signal. The data does not demand initial temperatures beyond those found  a moderate range: 200 $\leq T_{i} \leq$ 240 MeV.  Even though there might be still some unavoidable uncertainties in theoretical analyses, this conclusion appears robust.

\section{Photons from nuclear collisions at RHIC energies}
\subsection{A theoretical interlude on jet quenching}

\begin{figure}[htb]
\begin{center}
\includegraphics[angle=0,width=7.5cm]{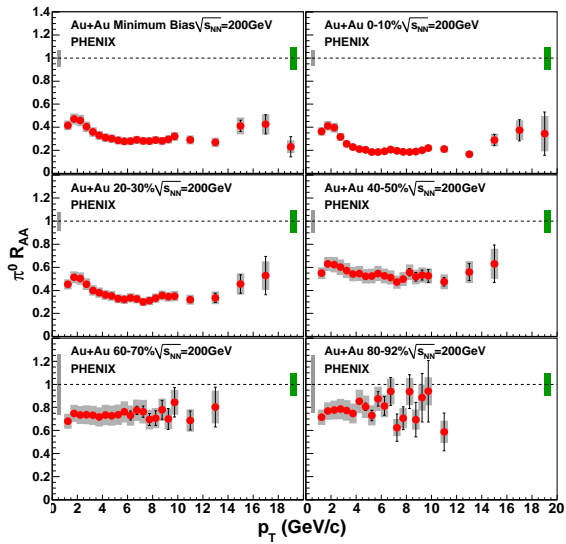}
\end{center}
\caption{The nuclear modification factor, $R_{AA}$, for $\pi^{0}$, measured by the PHENIX Collaboration in Au + Au collisions at RHIC. The different panels (left to right, top to bottom) start by a minimum bias sample and proceed through increasing centrality classes or, alternatively, from central to peripheral collisions. The figure is from  Ref. \cite{phenix-raa}.}
\label{PHENIX_RAA}
\end{figure}

The advent of RHIC - with an order of magnitude higher centre-of-mass energy than what was accessible at the SPS - has presented the community with incomparable opportunities for the formation and the study of the quark-gluon plasma. This entire program has been an unqualified success, which continues to this day. Arguably one of the main discoveries at RHIC has been the measurement of the spectacular absorption of the high energy jets by the strongly interacting matter \cite{jet-expt,dent}. In this context, it had been proposed early on that elastic final-state energy loss of fast partons would suppress jets in high-energy proton-proton events \cite{BJ}. Even though the medium created in nucleon-nucleon interactions turned out to be an inefficient sink for hard partons, subsequent studies following this seminal work have suggested that medium-induced radiative energy loss could in fact dominate its elastic counterpart, and that ``jet quenching'' could be seen in nucleus-nucleus collisions \cite{guy-jet}. A variable which characterizes this behavior and which uses the data in proton-proton collisions as a baseline is the nuclear modification factor, for a given nucleus-nucleus process (A+B) in a given class of centrality related to an impact parameter $b$:
\bea
R_{AA} (y, p_{T}, b) = \frac{d^{3}N^{AB}/dy d^{2} p_{T}}{\langle N_{\rm coll} (b)\rangle d^{3} N^{NN}/dy d^{2}p_{T}}
\label{raa}
\eea
where $d^{3}N^{AB}/dy d^{2}p_{T}$ is the spectrum of a given species, say, measured in nucleus-nucleus events (A+B). The denominator of Eq. (\ref{raa}) is the same spectrum measured in pp collisions, scaled by the number of inelastic collisions. Thus by design, if the nuclear process is only a superposition of what happens in  proton-proton events, $R_{AA} = 1$. The nuclear modification factor has been measured by different experimental collaborations, for many different hadron species,  at RHIC. Results obtained by PHENIX for neutral pions are shown in Figure \ref{PHENIX_RAA}. 
These results are striking. At such values of transverse momentum, the yield is naively expected to be dominated by QCD jet fragmentation. If the jets have the same characteristics as they do in vacuum, then the nuclear enhancement factor would be unity. It clearly is not. In addition, all experiments involving nuclei performed prior to the RHIC era have shown enhancement not suppression: As mentioned previously this has been ascribed to the Cronin effect. This experimental fact has recently been studied in Ref. \cite{phenix-cro}, and the different theoretical models used in this context were discussed and compared in Ref. \cite{acc}. The current status of our understanding of the results shown in Figure \ref{PHENIX_RAA} is that the exiting jets lose energy owing to their strong interaction with the surrounding medium created in the relativistic nuclear collisions. Using the framework of perturbative QCD, the energy loss by gluon radiation induced by interactions with the hot medium has been found to be the dominant mechanism \cite{eloss}. 
%However, it is becoming clearer that collisional energy loss can not, and in fact should not, be neglected entirely \cite{ecoll}. 
To quantify these statements, we continue to use the language of field theory at finite temperature and use the formalism of AMY (Arnold, Moore, and Yaffe), as applied to jet energy loss \cite{Jeon:2003gi,Turbide:2005fk}.

In this framework, the hard gluon and hard quark (or antiquark) distributions, $P_g$ and $P_{q}$, evolve in time as they traverse the medium. The evolutions are coupled, and proceed through \cite{Jeon:2003gi,Turbide:2005fk}
\bea 
\frac{dP_q (p)}{dt} \!\!& = &\!\! \int_k \! P_q (p+k) \frac{d\Gamma^q_{\!qg}(p+k,k)}{dkdt} - P_q
(p)\frac{d\Gamma^q_{\!qg}(p,k)}{dkdt} + 2 P_g (p+k)\frac{d\Gamma^g_{\!q \bar {q}}(p+k,k)}{dkdt}
\, , \nonumber \\
\frac{dP_g (p)}{dt} \!\!& = &\!\! \int_k \! P_q (p+k) \frac{d\Gamma^q_{\!qg}(p+k,p)}{dkdt} + P_g
(p+k)\frac{d\Gamma^g_{\!gg}(p+k,k)}{dkdt} \nonumber \\ &&\! \; - P_g (p) \left(\frac{d\Gamma^g_{\!q \bar
{q}}(p,k)}{dkdt} + \frac{d\Gamma^g_{\!gg}(p,k)}{dkdt} \Theta(2k-p) \!\!\right) \label{FP_eq}
\eea
where the integrals are over $k$ and run from $-\infty$ to $\infty$. The integration range with $k<0$ represents absorption of
thermal gluons from the QGP; the range with $k>p$ represents annihilation against an anti-quark from the QGP, of
energy $(k{-}p)$. The quantities $d\Gamma/dk dt$ are the transition rates for the various processes ($q~\to~q g$, $g~\to~q \bar{q}$, $g~\to~gg$), and be obtained by taking the imaginary part of the parton self-energy in the soft thermal background. Technically, an integral equation can be written and solved \cite{Jeon:2003gi,Turbide:2005fk}. In writing Eq.~(\ref{FP_eq}), we used $d\Gamma^g_{\!gg}(p,k)=d\Gamma^g_{\!gg}(p,p{-}k)$ and
similarly for $g\rightarrow qq$; the $\Theta$ function in the loss term for $g \rightarrow gg$ prevents double
counting of final states. 
\begin{figure}[htb]
\begin{center}
\includegraphics[angle=0,width=8cm]{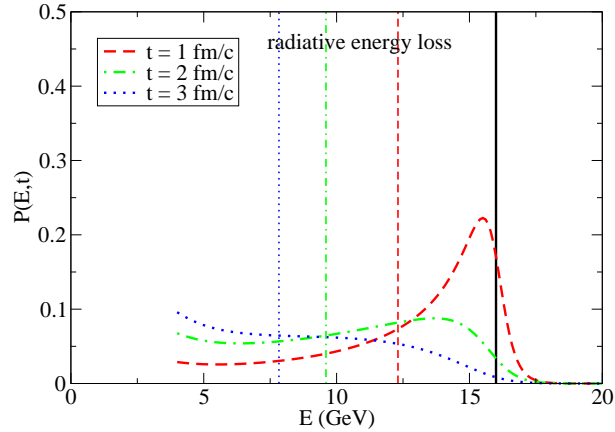}
\end{center}
\caption{  The evolution of a quark jet with initial energy $E_i = 16~{\rm GeV}$ propagating through a medium of
temperature $T = 400~{\rm MeV}$, where the vertical lines represent the values of mean energies related to the
corresponding distributions. The figure is from \cite{QinPhD}. \label{single_radiative}}
\end{figure}
The effect of the radiative energy loss on the parton distribution of an initially monoenergetic quark jet is shown on Fig. \ref{single_radiative}. The vertical lines show the average energies at different times. This plot shows how efficient the hot static (and here infinite) medium is in knocking apart the initial delta function. It also shows how rapidly the initial symmetrical profile will get distorted: It is important to work with the whole distribution, not just with the average energy. 

\begin{figure}[htb]
\begin{center}
\includegraphics[angle=-90,width=9cm]{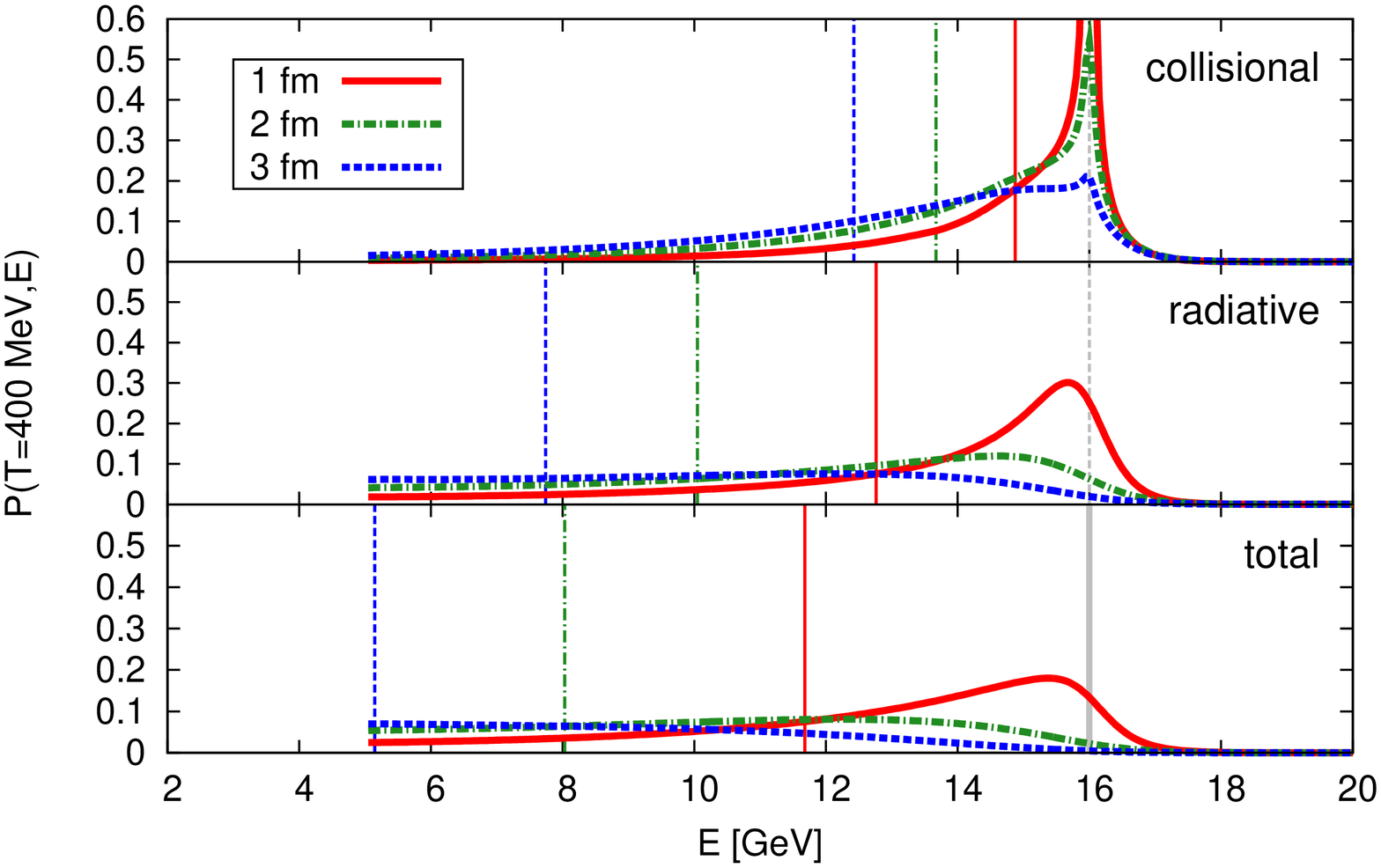}
\end{center}
\caption{  The evolution of a quark jet with initial energy $E_i = 16~{\rm GeV}$ propagating through a static medium of
temperature $T = 400~{\rm MeV}$, where the vertical lines represent the values of mean energies related to the
corresponding distributions. The top panel shows the effect of including only energy loss (and gain) through elastic collisions. The middle panel shows the effect of including only the  radiative energy loss and gain. The bottom panel includes both. The figure is from the model of Ref. \cite{schenke} and is courtesy of Bj\"orn Schenke. \label{elastic}}
\end{figure}
In addition to the energy loss (and possible gain) by the radiative emission of gluons (and also possibly absorbing them), collisional energy loss is also possible. This is often termed ``elastic energy-loss'': A slight misnomer as the total energy is of course conserved during an elastic collision, but we are concerned here only with the energy of a subset of the participants. A study of the qualitative modifications of the energy profile brought about by the inclusion of a collisional component is instructive. This exercise was done with an approximate treatment for the elastic energy loss, where only the drag and diffusion terms were kept \cite{QinPRL}. We show here results with a more complete treatment of the collisional loss (and gain) \cite{schenke}: This is done in Figure \ref{elastic}. It is important to reemphasize again  that any observables that will depend on the details of the energy profile will in principle be sensitive to the details being shown in Figure \ref{elastic}; a detailed analysis which includes those different aspects is being completed. It is however becoming clear that one should not neglect collisional energy loss. 

A complete study of the phenomenology associated with the jet energy loss scheme requires:
\begin{itemize}
\item Modeling the initial-state geometry of the strongly interacting zone  
\item Modeling the space-time evolution of this system 
\item Modeling the initial distribution of hard partons
\end{itemize}
Given these elements, the jets are then evolved using Eq. (\ref{FP_eq}) against the changing background provided by the evolution model. There are several ways the individual elements in the list above can be realized. An important constraint is provided by the need to reproduce as many complementary physical observables as possible; this again is certainly one of the lessons learned studying the physics  of relativistic nuclear collisions. This statement actually highlights the connection between the physics of jet quenching and of photon emission:
\begin{enumerate}
\item Hard quarks that lose energy to the medium and that will hadronize into jets will have a fragmentation component into photons and dileptons
\item Jet quenching is a jet-medium interaction. This class of interactions also has the potential to generate photons, real and virtual
\end{enumerate}
\begin{figure}[hbt]
\begin{center}
\includegraphics[angle=0,width=9cm]{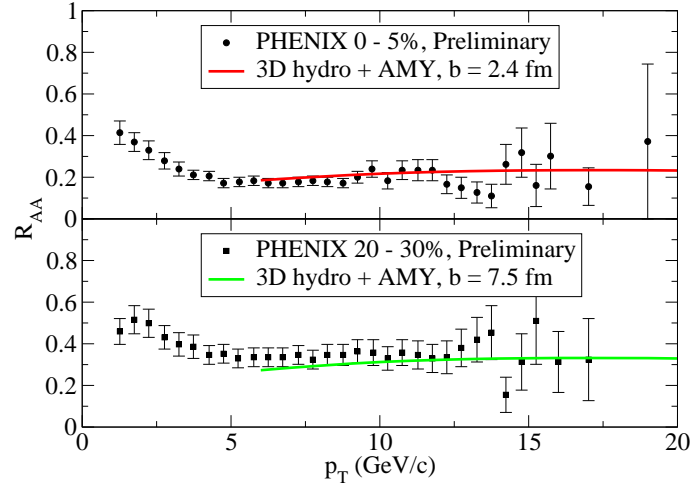}
\end{center}
\caption{The nuclear modification factor for measured $\pi^{0}$s, shown for central collisions (upper panel), and for midperipheral collisions (lower panel). The evolution model is an ideal (non viscous) 3D hydrodynamics \cite{nobass}. The figure is from Ref. \cite{QinPhD}, the data from Ref. \cite{phenix}. \label{RAA-pi0}}
\end{figure}
Thus, hard jet quenching and photon production at moderate to high $p_{T}$ can not be treated separately any longer. Several of the approaches used to calculate the energy loss of hard jets to the strongly interacting medium can be used to illustrate the connection between photons and hadrons born out of jet fragmentation; a comparative of several of these formalisms was recently done \cite{Bass:2008rv} in heavy ion collisions, where the time evolution model was the same in all cases.

In AMY, an overall free parameter is $\alpha_{s} = g^{2}/4\pi$, where $g$ is the strong interaction coupling constant. It is convenient to fit this to hadronic observables, like hadronic spectra or $R_{AA}$. An example of such a fit is shown in Figure \ref{RAA-pi0}, where the energy loss has been kept of the radiative type, and where the collision zone undergoes 3D hydrodynamical evolution. Note that the experimental nuclear modification factor for neutral pions is correctly reproduced in central and midperipheral collisions with $\alpha_{s} = 0.33$. Naturally, keeping $\alpha_{s}$ constant is an approximation and this value should be regarded as an average effective coupling. Note that, in the calculation of $R_{AA}$, the single $\pi^{0}$ distribution first needs to be correctly reproduced \cite{TurPhD,QinPhD}. Note also that in a three-dimentional expanding medium, the transition rates in Eq. (\ref{FP_eq}) are first evaluated in the local rest frame of the fluid cell, then boosted to the frame where the cell has a flow parameter \bm{ $\beta$}.  Specifically,
\bea
\left. \frac{d \Gamma (p, k)}{dk dt}\right|_{\rm lab} = (1 - \bm{v_{\rm jet}}\cdot \bm{\beta}) \left. \frac{d \Gamma (p_{0}, k_{0})}{dk_{0} dt_{0}}\right|_{\rm local}
\eea
where \bm{$v_{\rm jet}$} is the jet velocity, $k_{0}$ and $t_{0}$ are the local momentum and proper time, respectively. 

With some quantitative control over the energy loss of hard partons and over the time-evolution of their distributions, one may now go back to the evaluation of photon spectra.

\subsection{Photons from Jet-Medium Interactions}
Recently, the in-medium conversion of a hard parton to a photon has been suggested as a new source of electromagnetic radiation \cite{FMS}.    Specifically, when a fast quark, antiquark or gluon traverses a hot QCD medium it may undergo an annihilation ($q + \bar{q} \to g + \gamma$) or Compton ($q + g \to q + \gamma$) process with a thermal parton. The production rate of photons is then
\bea
\frac{dR}{dy d^2 p_T} = \sum_f \left(\frac{e_f}{e}\right)^2 \frac{T^2 \alpha \alpha_s}{8 \pi^2} \left(f_q (p_\gamma) + f_{\bar{q}} (p_\gamma) \right)\left[ 2 \ln \left(\frac{4 E_\gamma T}{m^2}\right) - C_A - C_C \right]
\eea
where $C_A$ and $C_C$ are two numerical constants respectively related to the annihilation and Compton channel; $C_A = 1.916$ and $C_C = 0.416$. As the jets will lose energy during their crossing of the hot and dense strongly interacting medium, it is vital to consider this energy loss in order to properly assess the importance of this contribution. 

%\subsubsection{Bremsstrahlung Photons}
As hard partons crossing the medium can emit gluons and lose energy, they may also produce bremsstrahlung photons as they interact with the parton constituents. The bremsstrahlung rate is obtained by appropriately modifying that for gluon emission \cite{Turbide:2005fk}. The photon distribution associated with the bremsstrahlung process is therefore given by
\bea
\frac{d P_{\gamma} (p, t)}{d t} = \int dk\ P_{q \bar{q}} (p+k) \frac{d \Gamma^{q \to q \gamma} (p+k, p)}{dk dt}
\eea
Note that even though the photon energy can be high, its production rate is small. Photon emission can thus be calculated perturbatively, without energy loss of the fast parton to the actual photons. Convolving the photon distribution with the initial spatial distribution of jets will yield the final spectrum of bremsstrahlung photons.

\subsection{Non-Thermal Sources}
A source of photons that has no thermal component is that of ``prompt photons'': Radiation associated with primary nucleon-nucleon interactions. In a heavy ion collision, these will contribute in the very first instants of the scattering process, they constitute an irreducible background to the photons associated with the thermal processes one hopes to study. Therefore, nucleon-nucleon collisions constitute an important reference and their understanding thus forms a baseline requirement for theory. 

The prompt photon production in a proton-proton collision is given by
\bea
E \frac{d^{3} \sigma_{\rm prompt}^{pp}}{d^{3} p_{\gamma}} &=& \sum_{a, b} \int \ d x_{a} d x_{b} g(x_{a}, Q) g(x_{b}, Q) \left\{K_{\gamma} {\frac{d \sigma}{d t}}^{a  b \to c  \gamma} \frac{x_{a} x_{b}}{\pi ( x_{a} - e^{y} p_{T}/\sqrt{s})}\right. \nonumber \\
& & \times \delta \left(x_{b} - \frac{x_{a} p_{T} e^{- y}}{x_{a} \sqrt{s} - p_{T} e^{y}}\right)\nonumber \\
&& + \left. K_{\rm frag} {\frac{d \sigma}{d t}}^{a  b \to c  d} \frac{1}{\pi z} D_{\gamma/c} (z,Q) \right\}
\eea
\begin{figure}[htb]
\vspace*{0.5cm}
\begin{center}
\includegraphics[angle=-90,width=8.0cm]{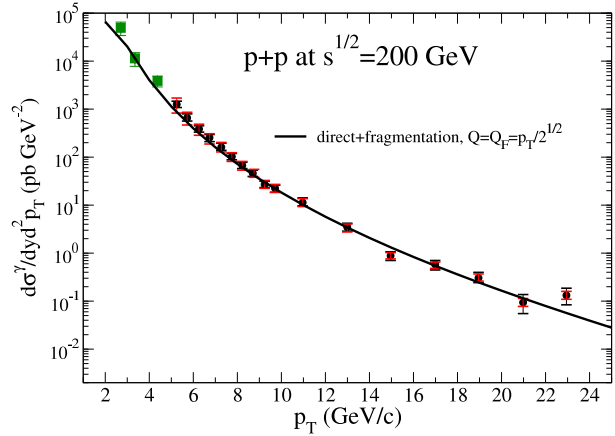} 
\end{center}
\caption{The spectrum of real photons measured in $p - p$ collisions at RHIC. The two sets of data points correspond to two diferent analysis techniques. The low $p_{T}$ data (green squares) is from Ref. \cite{:2008fqa}, the higher momentum data is from Ref. \cite{Adler:2006yt}. \label{ppphoton}}
\end{figure}
The different $K$-factors take into account NLO effects; they are evaluated here using the numerical program from Aurenche {\it et al.} \cite{patrick}. The photon fragmentation function, $D_{\gamma/c}$ is obtained from Ref. \cite{BFG}. An important purpose of this part of the photon production calculation is not only a reliable background estimate, but also a check of the NLO photon formalism at RHIC energies. Setting the factorization scale $Q$ equal to the fragmentation scale $Q_{F}$ and setting both equal to $p_{T}/\sqrt{2}$ yields the photon spectrum show in Figure \ref{ppphoton}. The high momentum data correspond to direct photon measurements, while the softer photons are actually lepton pairs that are treated as internal conversion photons. Note that it is crucial to get an accurate representation of the photon yields in pp collisions, as this constitutes the denominator of the nuclear modification factor $R_{AA}^{\gamma}$.
It is fair to write that the choice of scales yields a good agreement with the $p p$ data, and therefore constitute an appropriate prerequisite calibration for the calculations to follow.

\section{Photon yields at RHIC}
\subsection{1D Bjorken expansion}
Prior to using a more realistic scenario to be discussed below, in order to get some intuition it is useful to first benchmark the different photon sources using a simple time-evolution model of the nuclear collision. A 1D Bjorken expansion \cite{Bj} is therefore considered. Also, the initial temperature is assumed to scale with the local density, producing a temperature evolution of a QGP as
\bea
T(r, \tau) = T_{i} \left(\frac{\tau_{i}}{\tau}\right)^{1/3} \left[ 2 \left( 1 - \frac{r^{2}}{R_{\bot}^{2}}\right)\right]^{1/4}
\eea
The jet evolves in the QGP until it reaches the surface or until the temperature reaches its critical value, taken here to be $T_{c}$ = 160 MeV,  when a first-order phase transition occurs. The fraction of the QGP thereafter in the mixed phase is
\bea
f_{\rm QGP} = \frac{1}{r_{d} - 1} \left( \frac{r_{d} \tau_{f}}{\tau} - 1 \right)
\eea
where $r_{d}$ is the ratio of the number of  degrees of freedom in the two respective phases (partonic over hadronic), and $\tau_{f}$ is the time when the temperature reaches $T_{c} $. Recall that in a Bjorken expansion, the initial time and temperature are related by
\bea
T_{i}^{3} \tau_{i} = \frac{\pi^{2}}{\zeta (3) g_{Q}}\frac{1}{\pi R_{\bot}^{2}} \frac{d N}{d y}
\eea
where $g_{Q}$ is the number of degrees of freedom in the QGP phase. The measured particle rapidity density, $d N/d y$, thus determines the lifetime of the QGP. 

\begin{figure}[htb]
\begin{center}
\includegraphics[angle=-90,width=5.7cm]{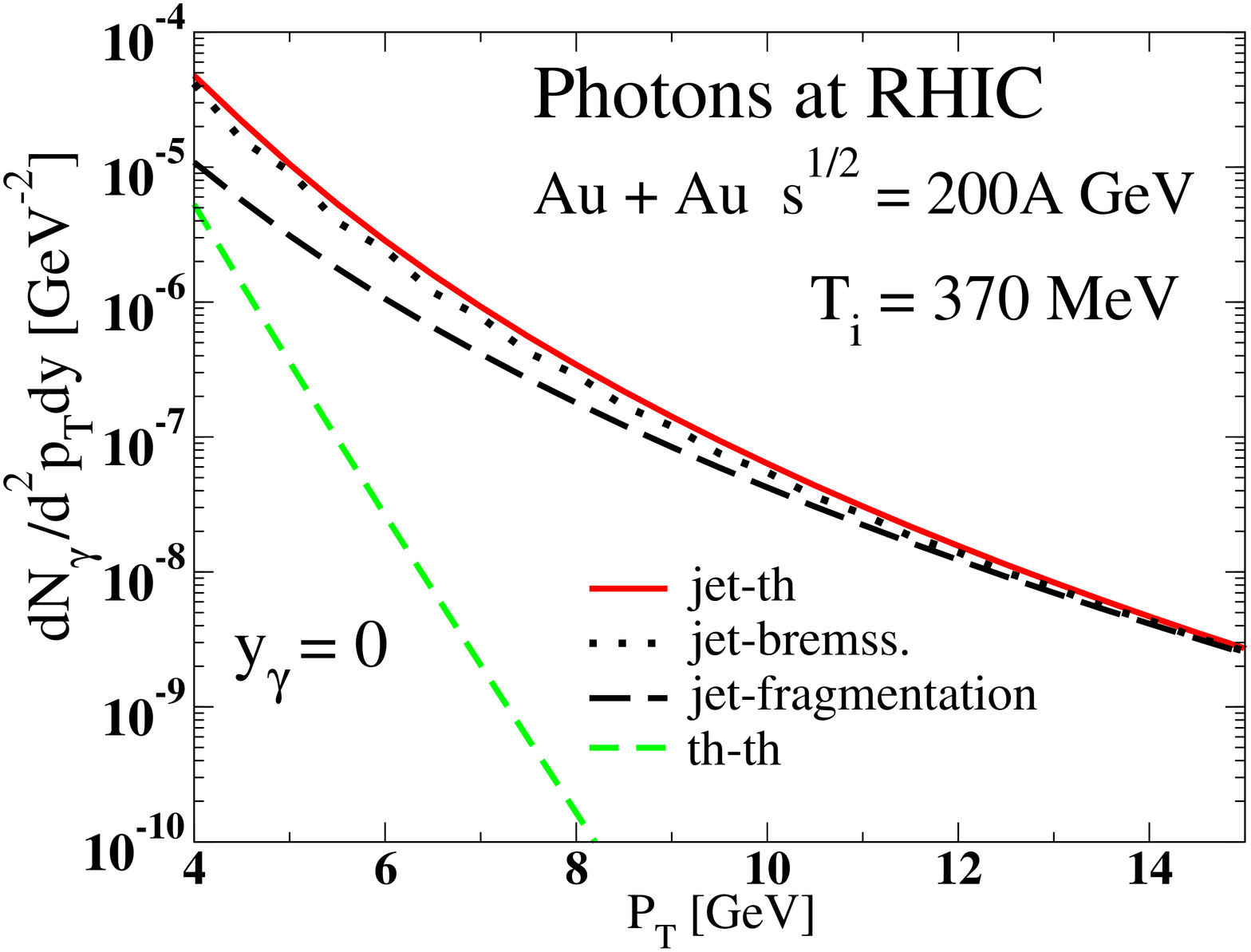}
\includegraphics[angle=-90,width=5.8cm]{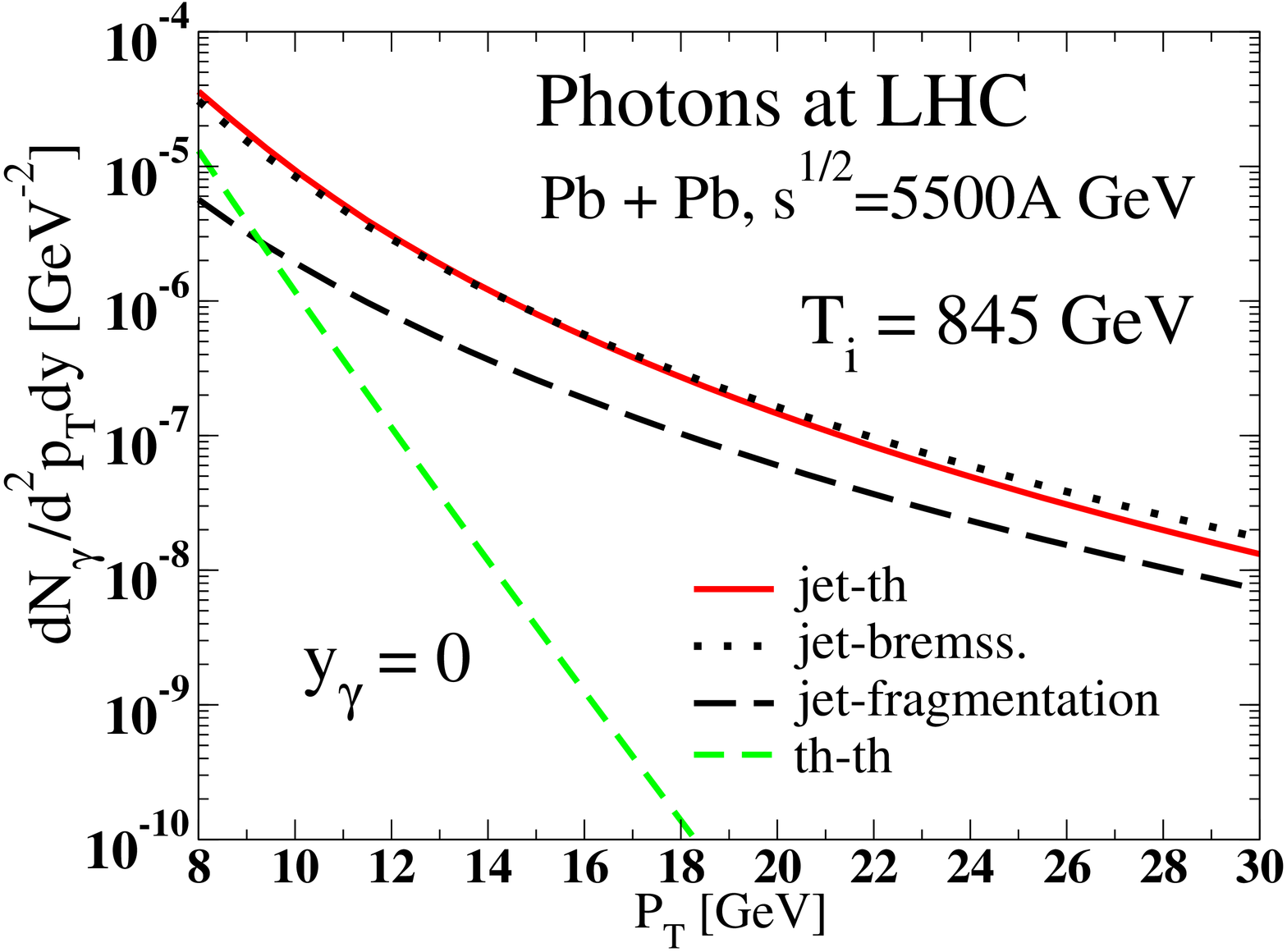}
\end{center}
\caption{Photon yields at the energies of RHIC (left panel) and of the LHC (right panel). The time-evolution model used here is a 1D Bjorken expansion. The difference sources shown are: Emission from a purely thermal source (th-th), photons emitted from a jet interacting with the thermal QGP (jet-th), bremsstrahlung photons (jet-bremss), and photons from a fragmenting jet that has exited the thermal environment. Note that the bremsstrahlung curves in Ref.  \cite{Turbide:2005fk} and those shown here are different \cite{err}. \label{BJ_yields}}
\end{figure}

A free parameter in the approach described so far is $\alpha_{s}$ (or equivalently the strong coupling constant, $g$). This is determined by requiring the measured nuclear modification factor for hadrons to be correctly reproduced, as discussed previously.
% That for neutral pions produced in symmetric collisions of gold nuclei at RHIC is shown in Figure \ref{R_aa}, and represent a value of $\alpha_{s} = 0.3$.
%\begin{figure}[htb]
%\begin{center}
%\includegraphics[width=8cm]{R_aa_new_pion.eps}
%\end{center}
%\caption{The nuclear modification factor for $\pi^{0}$'s produced in Pb - Pb collisions at $\sqrt{s}$ = 200 GeV/nucleon. The solid curve represent the model described in the text. \label{R_aa}}
%\end{figure}
After setting the value of the strong coupling constant, the photon production calculation may proceed without additional parameters. Integrating the rates introduced earlier with the 1D Bjorken model produces the yields displayed in Figure \ref{BJ_yields}. 

\begin{figure}[htb]
\begin{center}
{\includegraphics[width=5.0cm]{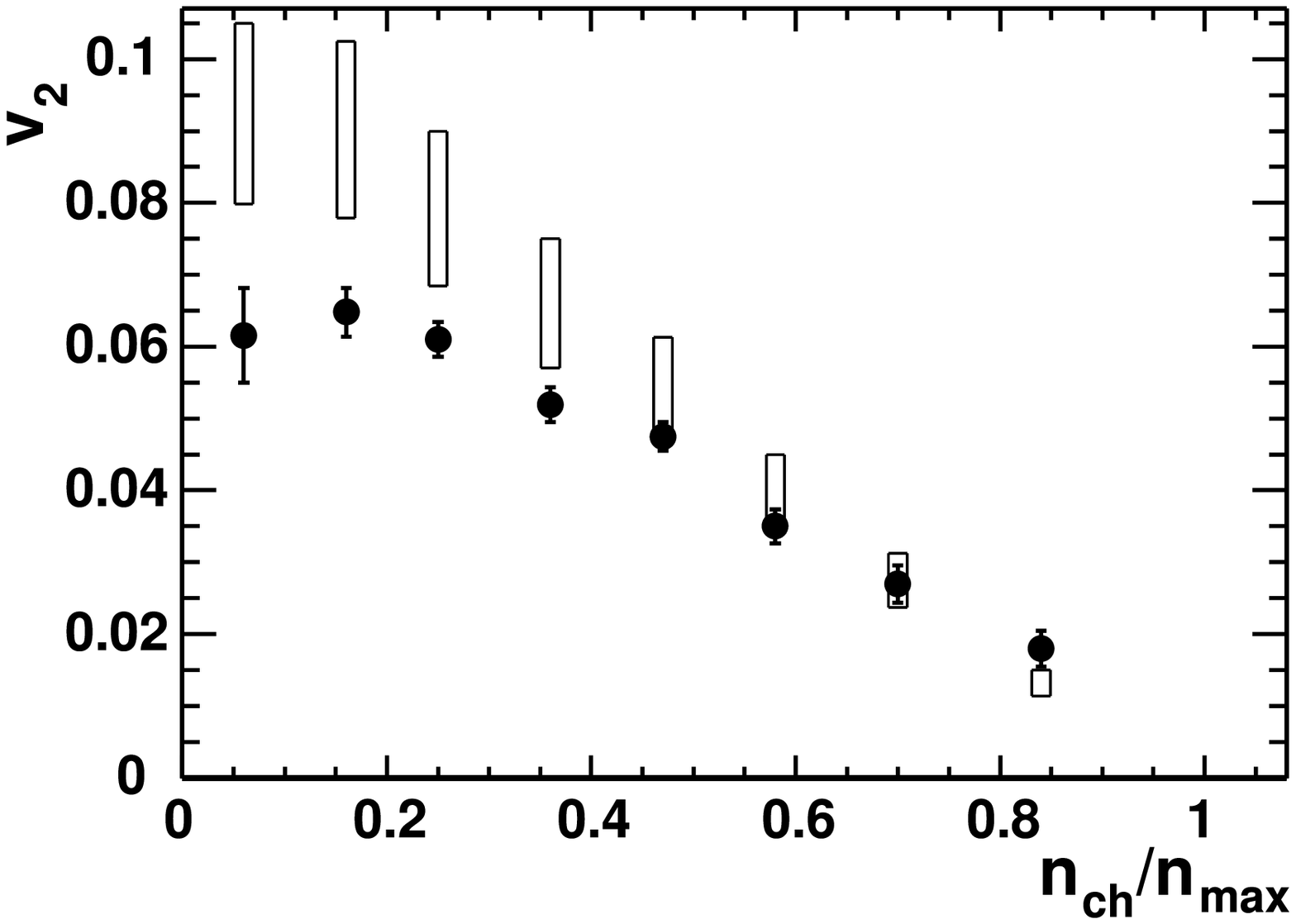} \hspace*{1cm} \includegraphics[width=4cm]{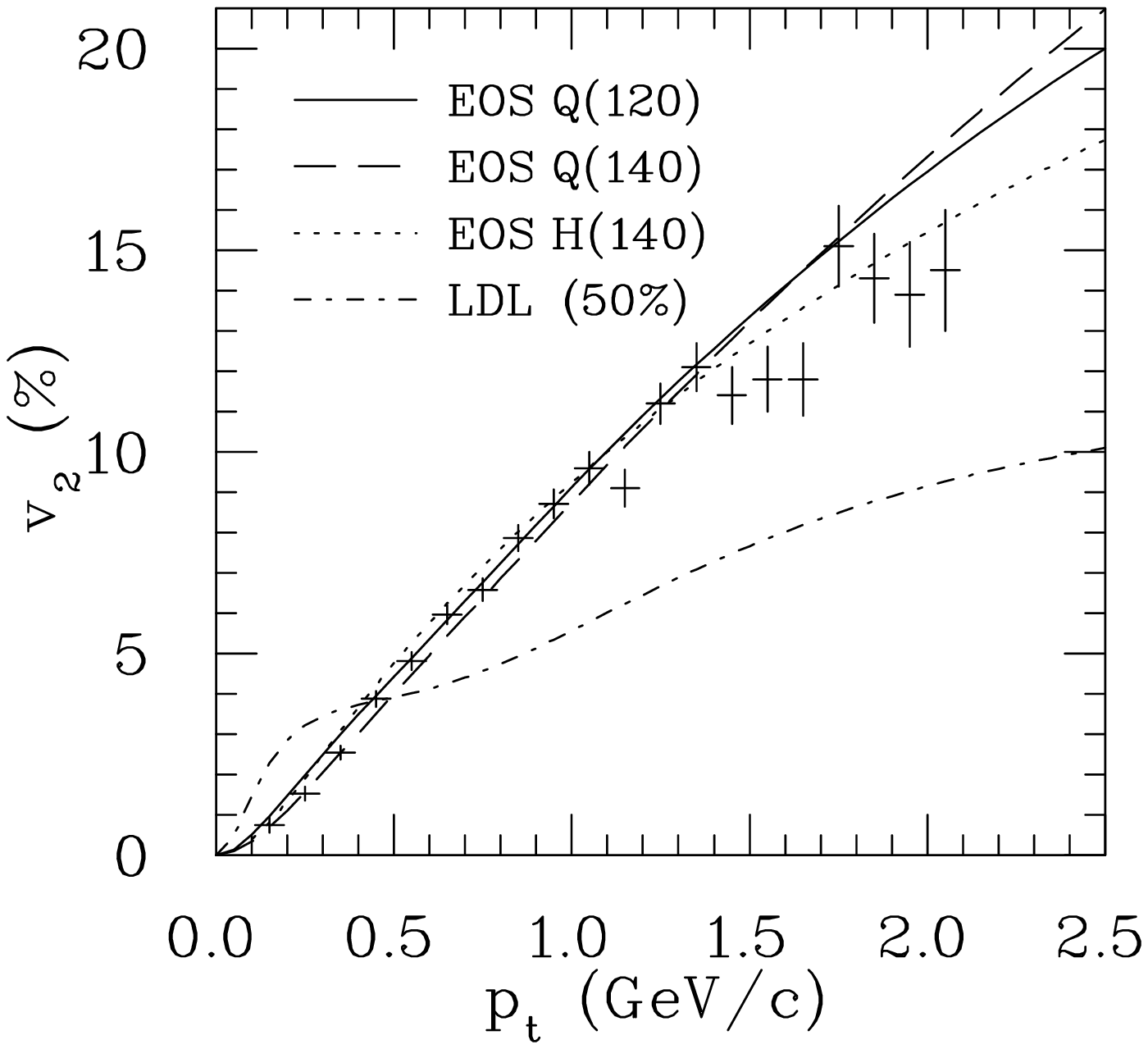}}
\end{center}
\caption{The elliptic flow of unidentified charged particles in Au + Au collisions, at 130 A GeV, as a function of the multiplicity of the events  (left panel). The vertical bars indicate the range of early hydrodynamical predictions. The right panel shows minimum bias data of the same nature, as a function of transverse momentum, $p_{T}$. The different solid curves represent hydrodynamic calculations with different equations of state and different freeze-out temperatures. From Ref. \cite{Kolb:2003dz}. \label{v2}} 
\end{figure}

It is clear that identifying different sources of photons simply from the signal strength represents a challenging tasks, as the size of the background contributions has to be known precisely. An experimental observable with a different and unique qualitative behavior would be a genuine asset. In this context, the azimuthal asymmetry of the electromagnetic signal was recently proposed as an observable with such potential. More precisely, a sizeable azimuthal asymmetry characterized by a coefficient $v_{2}$ can be expected for intermediate to large$p_{T}$ photons produced in non-central high energy nuclear collisions. It is helpful to recall that the anisotropy is defined in terms of the Fourier analysis of the particle yield $dN/p_{T} d p_{T} d \phi$, through the coefficients $v_{n}$:
\bea
\frac{dN}{p_{T} dp_{T} d\phi} = \frac{dN}{2 \pi p_{T} d p_{T}} \left[ 1 + \sum_{n} 2 v_{n} (p_{T}) \cos (n \phi)\right]
\eea
The angle $\phi$ is defined with respect to the reaction plane. At midrapidity, all the odd coefficients vanish by symmetry and $v_{2}$ is left as the dominant one. Importantly, at the energies of the SPS and of RHIC, hadrons at low and intermediate $p_{T}$ always have a positive $v_{2}$, owing to the hydrodynamic flow \cite{Voloshin:2008dg}. A good example is shown in Figure \ref{v2}.
On the other hand, propagating jets should lose more energy when they are sent into a region where the medium is thicker, namely out of the reaction plane. As hadrons at intermediate and high $p_{T}$ get an important yield contribution from jet fragmentation, a stronger jet quenching should lead to fewer hadrons with the appropriate value of their momentum emitted into this direction. Importantly this ``optical $v_{2}$'' is not associated with flow but with absorption; the argument implies a positive $v_{2}$ for hadrons originating from jets. No quenching would mean a vanishing azimuthal anisotropy for hadrons born from a fragmenting hard parton.  Measurements at RHIC, for various hadron species, support this interpretation \cite{RHIC_hadrons}. Concentrating on intermediate energy direct photons, it is clear from previous discussions that photons from jets (i.e. from fragmentation and from jet-medium interaction) will compete with those originating from primary hard scatterings of partons in the nuclei.

\begin{figure}[htb]
\vspace*{0.5cm}
\begin{center}
\includegraphics[width=7.8cm]{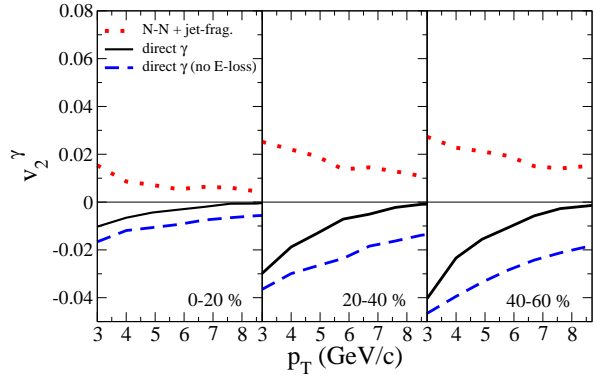} 
\end{center}
\caption{Photon $v_{2}$ as a function of transverse momentum, for three centrality classes in Au + Au collisions at the top RHIC energy. The dotted line shows $v_{2}$ for the photons originating from primary hard collisions and from the fragmentation of hard partons. The solid curve shows the anisotropy coefficient for photons from all sources. The dashed line shows the effect of considering all photons, but neglecting the energy loss of propagating jets. From \cite{TGF}.   \label{photonv2_1}} 
\end{figure}

To further quantify these arguments, a numerical calculation is done for Au + Au collisions at RHIC ($\sqrt{s_{NN}}$ = 200 GeV). Using again a simple 1D Bjorken expansion, the initial conditions are constrained by the measured particle densities. The overlap region of the two colliding contains the primordial hard partons which will hadronize after escaping. This initial almond-shaped profile is determined here using a simple hard sphere geometry.  Three centrality classes are considered: 0 - 20\%, 20 - 40\%, and 40 - 60\%; fixing the initial formation time at $\tau_{i}$ = 0.26 fm/c, the initial temperatures are respectively $T_{i}$ = 370, 360, and 310 MeV. The results for the photon azimuthal anisotropy coefficient are shown in Figure \ref{photonv2_1}.
Indeed, photons from primordial hard scatterings and from the fragmentation of hard jets show a positive vaue of $v_{2}$, owing to jet energy loss. Interestingly, the contribution from jet-plasma photons (in this transverse momentum window) is enough to make $v_{2}$ go negative. At lower values of $p_{T}$, photon anisotropy should receive a contribution from hydrodynamic flow, as do hadrons. 
\begin{figure}[htb]
\vspace*{0.5cm}
\begin{center}
\includegraphics[width=7.0cm]{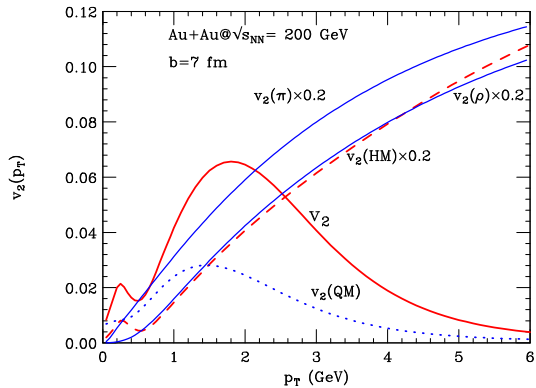} 
\end{center}
\caption{Photon $v_{2}$ as a function of transverse momentum, for Au + Au collisions at 200 $A$ GeV, for a fixed impact parameter of $b = $ 7 fm.  Quark and hadronic matter (dashed line) contributions  are shown separately, and the elliptic flow of $\pi$'s and $\rho$'s is shown for comparison. From Ref. \cite{Chatterjee:2005de} \label{Rupa_1}.}
\end{figure}
Such a hydrodynamic  calculation has been performed for low $p_{T}$ photons \cite{Chatterjee:2005de} and the results are shown in Figure \ref{Rupa_1}. Clearly, the $v_{2}$ coefficient for low $p_{T}$ photons has the potential to display a revealing transition from a behavior characteristic of hadronic matter, to one typical of partonic matter. A similar calculation exists for lepton pairs \cite{Chatterjee:2007xk} and thus lies outside the topics to be addressed here. Note finally that a complete picture demands a calculation where the soft and hard photons are reconciled in a common framework, and treated consistently within a given realistic dynamical model. The results of such a calculation are presented in the next section. 

\subsection{Photon yields and relativistic hydrodynamics}
Whenever computing electromagnetic observables in relativistic nuclear collisions, it is imperative to use a model whose parameters have not been specifically adjusted to fit those very observables, but can address a wide class of hadronic phenomena as well. It is therefore important to use ``realistic'' hydrodynamical simulations approaches, namely ones that reproduce many of the soft hadronic characteristics measured at RHIC. One such model is AZHYDRO. This ideal (i.e. where the shear and bulk viscosity coefficients are zero) hydrodynamic simulation in 2+1D  has been empirically successful at RHIC: It is able to reproduce the bulk of the data on soft hadron production for $p_{T} \leq$ 1.5 - 2 GeV/c. 
This is the case for hadronic momentum spectra from central to semiperipheral Au + Au collisions, including the azimuthal anisotropies \cite{Kolb:2003dz}. Apart from ideal hydrodynamics, AZHYDRO assumes a boost-invariant longitudinal flow velocity. For variables measured near midrapidity, this theoretical input is found to be appropriate. Going beyond this requires fully 3D hydrodynamic modeling. The different photon sources already introduced earlier are revisited briefly here, in the context of the hydro. 

For the photons resulting from jet-induced interactions, an important element is the initial momentum distribution of jets, at impact parameter $b$. This is obtained from
\bea 
\frac{d N_{\rm jet} (Q, b)}{d^{2} q_{T} dy} = T_{AB} (b) \sum_{i, j, k} \int d x_{i} g_{i} (x_{i}, Q) g_{j} (x_{j}, Q)\, K {\frac{d \sigma}{dt}}^{i +j \to k + {\rm jet}} \frac{ x_{i} x_{j}}{\pi  (x_{i} - e^{y} q_{T}/ \sqrt{s}) }\nonumber \\
\eea
The $K$-factor includes and accounts for next-to-leading order effects. Here, a value of $K$ = 1.7 will be used, which is essentially transverse momentum independent \cite{bfl}. Importantly, as will be seen later, isospin effects are included at the parton level:
\bea
g_{a} (x_{a}, Q) = \left( \frac{Z}{A} f_{a} (x_{a}, Q) + \frac{A-Z}{A} f_{a*} (x_{a}, Q) \right) R(x_{a}, Q)
\eea
where $f_{a}$ is the parton distribution function inside a proton. The second term, $f_{a*}$, corresponds to the distribution function inside a neutron and is obtained by the appropriate substitution of parton species.  Shadowing effects are included in the function $R( x_a, Q)$ \cite{kari}. The factorization scale $Q$ is taken as $q_{T}$.

The full predictive power of the relativistic hydrodynamic is now invoked, in connection with the evolution of the hard partons characteristics. 
%In the laboratory frame (that of the nucleus-nucleus collision) the transitions rates are
%\bea
%\frac{d \Gamma_{q g}^{q}}{d k dt} = (1 - {\bf v}_{\rm jet} \cdot {\bf \beta}) \frac{d \Gamma_{q g}^{q}}{d k_{0} dt_{0}}
%\eea
%where $d \Gamma_{q g}^{q} / d k_{0} dt_{0}$ is evaluated in the rest frame of the fluid. The local fluid cell is moving with a velocity ${\bf \beta}$ in the laboratory frame; the two frames are related by the Jacobian $(1 - {\bf v}_{\rm jet} \cdot {\bf \beta})$. 
The local transition rates in each fluid cell are taken from the solution of Eq. (\ref{FP_eq}), and these depend on the strong coupling constant $\alpha_{s}$, and on the local temperature $T$. This quantity is evolved in he longitudinally boost-invariant AZHYDRO with the EOS Q equation of state \cite{Kolb:2003dz} which matches a non-interacting QGP above $T_{c}$ to a chemically equilibrated hadron resonance gas below $T_{c}$ at a critical temperature of $T_{c}$ = 164 MeV. A somewhat early thermalization, $\tau_{i}$ = 0.2 fm/c is assumed, to partially account for a pre-equilibrium stage \cite{Chatterjee:2005de}. Further note that the transverse flow velocity at any vale of the space-time rapidity can be inferred from that at $\eta = 0$. Writing $\bm{\beta} = (\bm{ \beta_{\bot}}, \beta_{z})$, we have
\bea
\bm{ \beta_{\bot}} (\tau, \eta, {\bf x_{\bot}}) = \frac{\bm{ \beta_{\bot}} (\tau, \eta = 0, {\bf x_{\bot}})}{\cosh \eta}, \ \ \beta_z = (\tau, \eta, {\bf x_{\bot}}) = \tanh \eta
\eea

\begin{figure}[htb]
\begin{center}
\includegraphics[angle=-90,width=6.0cm]{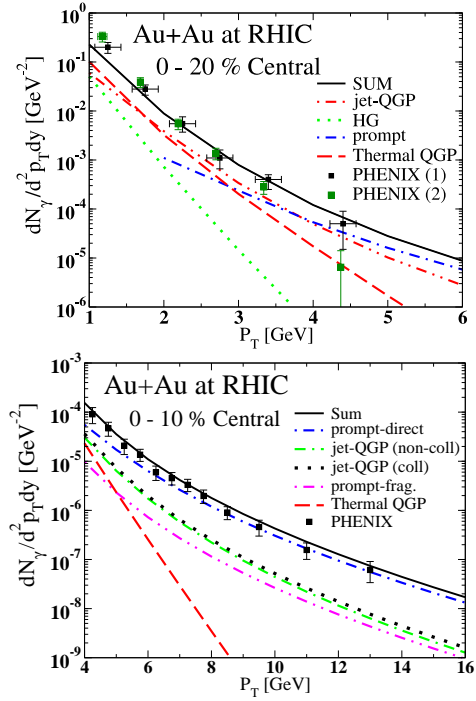} \\ 
\includegraphics[width=6.2cm]{photon_010_2D.eps}
\end{center}
\caption{The spectrum of real photons measured in Au - Au  collisions at RHIC. The top panel data is extracted following the same technique (identifying low mass dileptons with a virtual photon) as that used for the low momentum part of Figure \ref{ppphoton}, and is for a centrality class of 0 - 20\%. The data set ``PHENIX (1)'' is from \cite{:2008fqa}, while the data set ``PHENIX (2)'' is from \cite{busch}. The latter supersedes the former. The bottom panel is for a centrality class of 0 - 10\%; the higher momentum data there corresponds to a direct measurement and is from Ref. \cite{isobe}. The different contributions are discussed in the main text.
\label{Auphotons}}
\end{figure}

Integrating the rates with the time-evolution generated by the relativistic hydrodynamics produces observables one can then compare with experimental measurements. The hadronic variables produced by AZHYDRO are discussed elsewhere \cite{Kolb:2003dz}, and the photon spectra are shown in Figure \ref{Auphotons}. 
The contributions from hard primordial scatterings are labeled ``prompt'' in the top panel and include photons from Compton and annihilation events, together with those from jet fragmentation. The radiation from the thermal components of the quark-guon plasma is shown (``Thermal QGP''), and so is that from a hot gas of hadrons (``HG''). Comparing the sum of the different contributions (the solid curve) with the experimental data, it appears that the photons originating from jet-plasma interactions constitute a non-negligible portion of the signal in the momentum window 2 $< p_{T} <$ 4 GeV. The lower momentum data are dominated by thermal emission, while the radiation from hard collisions gradually takes over the whole spectrum,  for growing values of transverse momentum.  Most of the higher momentum data in the lower panel can be attributed to Compton and annihilation contributions calculated from pQCD. The purely thermal contributions are subdominant there and the fragmentation contribution to the real spectrum is small, owing mainly to the energy lost by the propagating jet. Adding together all the sources produces a good agreement with the data, with no additional adjustment of the  model parameters (hydro characteristics, $\alpha_{s}$, \ldots). 

\begin{figure}[htb]
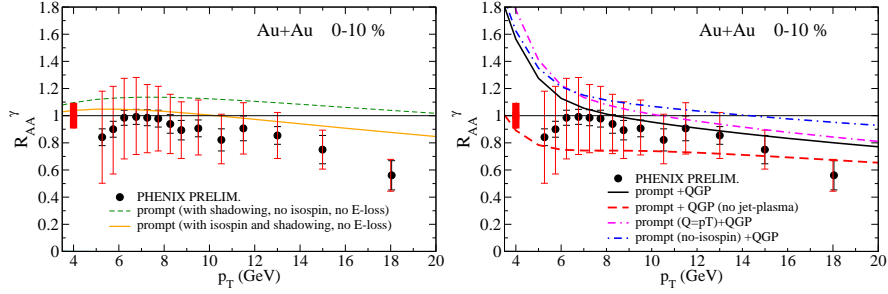

\begin{center}
\includegraphics[width=5.8cm]{R_AA_phot_1.eps} 
\includegraphics[width=5.8cm]{Raa_new_photon.eps} 
\end{center}
\caption{The nuclear modification factor for direct photons produced in central Au + Au collisions at RHIC, calculated using 2 + 1D hydro. In the pQCD contribution, the scale used is $Q = \sqrt{p_{T}}/2$. The left panel contains no finite temperature effects and shows the importance of shadowing and and of accounting for isospin. In the right panel, the effect of a QGP and of varying the scale are considered. The dashed line shows  results obtained without considering the photons originating from jet-medium interactions. From Ref. \cite{TGFH}.    \label{RAA}}
\end{figure}

Another useful representation of the same data is the nuclear modification factor now defined for real photons as
\bea
R_{AA}^{\gamma} (b, p_{T}, y) = \frac{\int_{0}^{2 \pi} d \phi d N^{\gamma} (b)/d^{2} p_{T} dy}{2 \pi T_{AB} (b) d \sigma_{\rm prompt}^{pp} / d^{2}p_{T} dy}
\eea
we only consider $y = 0$ in this work. Also, as advertised previously, the azimuthal anisotropy coefficient might help disentangle some of the photon sources. Both these projections of the data are examined. In what concerns $R_{AA}^{\gamma}$, it is first useful to isolate some of the cold nuclear matter effects; this is done in the left panel of Figure \ref{RAA}.
In these estimates, a considerable effect on the nuclear modification factor is caused by neglecting the jet-plasma photons. This amounts to a reduction of approximately 30\% (at intermediate values of $p_{T}$), as seen in the right panel of Fig. \ref{RAA}. The two extreme cases - where jet-plasma photons are present or not - bracket the experimental data; the current large error bars do not permit a choice. The apparent downward trend of the data is intriguing. Isospin contributes to this as noticed in Ref. \cite{Arleo:2006xb}, and seen in the left panel. Notably, in the calculations presented here, the additional suppression in $R_{AA}^{\gamma}$ originates from the fact that jets fragmenting into photons have lost energy. This constitutes an important conceptual link between hadronic physics and electromagnetic observables. The large values of the nuclear modification factor at low momenta $p_{T} < $ 6 GeV can be directly related to thermally-induced channels, and appear to exceed somewhat the central values of the experimental measurements: Note however that pQCD slightly underestimates the denominator in $R_{AA}^{\gamma}$. Correcting this would help bring the calculation closer to what is determined experimentally on this linear plot. Smaller error bars will also enable progress in identifying the respective sources, as well as eventual contributions from cold nuclear matter \cite{cold}. 

\begin{figure}[htb]
\begin{center}
\includegraphics[width=7.0cm]{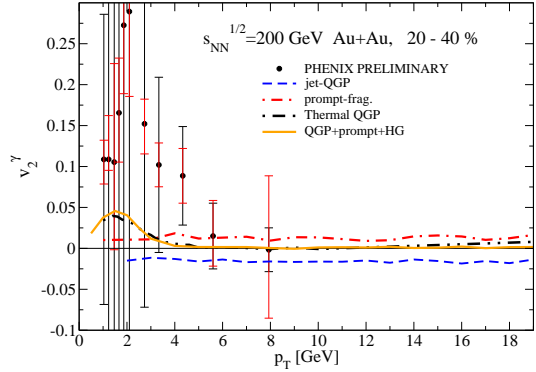} 
\end{center}
\caption{The nuclear modification factor for direct photons produced in central Au + Au collisions at RHIC, calculated using 2 + 1D hydro. In the pQCD contribution, the scale used is $Q = \sqrt{p_{T}}/2$. The left panel contains no finite temperature effects and shows the importance of shadowing and and of accounting for isospin. In the right panel, the effect of a QGP and of varying the scale are considered. The dashed line shows  results obtained without considering the photons originating from jet-medium interactions. From Ref. \cite{TGFH}.    \label{photonv2}}
\end{figure}

Turning to the determination of photon azimuthal anisotropy, results of hydrodynamical simulations can also be compared with experimental data. Hydro calculations for low $p_{T}$ photons were shown in Fig. \ref{Rupa_1}, and Fig. \ref{photonv2_1} displays results for higher $p_{T}$ which however used a Bjorken model expansion. Both regions are unified using a realistic evolution scenario  and a realistic initial profile. Considering 
\bea
v_{2}^{\gamma} (b, p_{T}, y) = \frac{\int_{0}^{2 \pi} d \phi \cos 2 \phi d^{3} N^{\gamma} (b) /d^{2} p_{T} d y}{\int_{0}^{2 \pi} d \phi d^{3}N^{\gamma}(b)  / d^{2} p_{T} d y}\ ,
\eea
the results of the theoretical calculation are compared with the experimental data in Figure \ref{photonv2}. 
One sees that the experimental anisotropy is small and in fact essentially zero for $p_{T} > $ 5 GeV. The $v_{2}$ coefficient associated with jet-plasma interactions is indeed negative, but its influence is greatly diminished as compared with earlier estimates of Figure \ref{photonv2_1}, associated with simpler dynamics and geometry. Future analyses with isolation cuts for the jets might however restore this sensitivity.

\section{Tagging jets with electromagnetic radiation}
\label{sec:2}

It has become clear that differential variables will offer higher discriminating power than their one-body counterpart, in what concerns the dynamics of relativistic nuclear collisions. The nuclear modification factor, $R_{AA}$, for example, offers only moderate sensitivity to the underlying probability distribution of energy loss \cite{Renk:2007mv}.   In this context, correlation measurements could go a long way in constraining energy loss schemes. The yield of associated hadrons, given a trigger hadron, is sensitive to details of the evolution model for di-hadron correlations, much more so than $R_{AA}$ \cite{Renk:2006pk,Majumder:2008jy}. Often termed the ``golden channel'', a very promising trigger of hard hadrons is a high $p_T$ photon \cite{photrig}. The argument can be succinctly summarized as follows: At leading order, direct photons are exactly correlated to a recoiling  parton, in a hard parton-parton collision such as one occurring in a Compton process or in an annihilation process.  Modulo next-to-leading-order effects, measuring a hard photon on the near side yields valuable information about the far-side energy of the recoiling jet. Measuring the fragmentation hadrons then will directly contain info pointing to the lost energy. 

\begin{figure}[htb]
\begin{center}
\includegraphics[width=7.0cm]{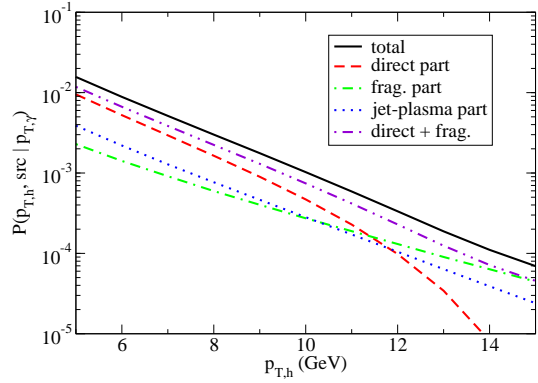} 
\end{center}
\caption{Given a 15 GeV photon as a trigger in central Au-Au collisions, the yield per trigger of away-side hadrons is shown and is further separated in different contributions discussed in the text.  From Ref. \cite{hp2008}.    \label{taghadrons}}
\end{figure}

An observable amenable to the study of correlations, and which contains more information that one-body distributions is the so-called yield per trigger, a conditional probability distribution:
\bea
P(p_{T}^{h} | p_{T}^{\gamma}) = \frac{P(p_{T}^{h}, p_{T}^{\gamma})}{P(p_{T}^{\gamma})}
\eea
In the above, $P(p_{T}^{\gamma})$ is the single photon distribution, $p_{T}^{h}$ is a hadron transverse momentum, and $P(p_{T}^{h}, p_{T}^{\gamma})$ is the two-particle (photon-hadron) distribution \cite{Qin2009}:
\bea
P(p_T^\gamma) \!&=&\! \int \frac{d\phi}{2\pi}\! \int \! d^2\vec{r}_\bot {\cal P}_{AB}(\vec{r}_\bot) \nonumber \\&& \times \!\sum_{j} \! \int \! dp_T^{j} P(p_T^j)
P(p_T^\gamma|p_T^j,\vec{r}_\bot, \phi),  \ \ \ \ \ \ 
\nonumber \\
P(p_T^h,p_T^\gamma) \! &=&\! \int \frac{d\phi}{2\pi} \! \int \! d^2\vec{r}_\bot {\cal P}_{AB}(\vec{r}_\bot)
\nonumber \\ && \times 
\!\sum_{jj'} \! \int \! dp_T^{j} dp_T^{j'} P(p_T^j, p_T^{j'})
\nonumber \\ && \times P(p_T^\gamma|p_T^j,\vec{r}_\bot, \phi) P(p_T^h|p_T^{j'},\vec{r}_\bot, \pi \!+\! \phi). \ \ \ \ \ \ 
\eea
Note that ${\cal P}_{AB} (r_{\bot})$ is the usual product of nuclear thickness functions, where the impact parameter dependence is implicit. Explicitly, ${\cal P}_{AB} (b, {\bf r_{\bot}}) = T_{A}({\bf r_{\bot}} + {\bf b}/2) T_{B} ({\bf r_{\bot}} - {\bf b}/2)/T_{AB} (b)$,
 and $T_{AB}(b) = \int d^{2} r_{\bot} T_{A} ({\bf r_{\bot}}) T_{B} ({\bf r_{\bot}} + {\bf b})$. The definition of the nuclear thickness function in terms of, say, a Woods-Saxon form is $T_{A} ({\bf r_{\bot}}) = \int dz\ \rho_{A} ({\bf r_{\bot}}, z)$. In addition, $P(p_T^j)$ and $P(p_T^j, p_T^{j'})$ are the probability distributions for single-jet and back-to-back di-jet production. Studies of photon-hadron correlations make it useful to define the fragmentation function in terms of the yield per trigger:
 \bea
 D_{AA} (z_{T}, p_{T}^{\gamma} ) = p_{T}^{\gamma} P (p_{T}^{h} | p_{T}^{\gamma})
 \eea
 where $z_{T} = p_{T}^{h} / p_{T}^{\gamma}$. Finally, we define another nuclear modification factor in terms of this photon-triggered fragmentation function, normalized to that in pp collisions:
\bea
  I_{AA} (z_{T} , p_{T}^{\gamma} ) = \frac{D_{AA} (z_{T}, p_{T}^{\gamma})}{D_{pp} (z_{T}, p_{T}^{\gamma})}
\eea
We may now look at experimental data, with a theoretical approach which includes
\begin{itemize}
\item All the photon sources discussed so far, but neglecting the purely thermal ones. We shall restrict our analyses to photon in the intermediate to hard sectors, where thermal emission can be neglected.
\item Both radiative and elastic energy loss (and gain) of the jets 
\item The space-time evolution of the strongly interacting system, as modeled by relativistic hydrodynamics
\end{itemize}
Before comparing with measurements, it is instructive to highlight the power of this theoretical tool by looking at variables that cannot  be directly measured. 
\begin{figure}[htb]
\begin{center}
\includegraphics[width=7.0cm]{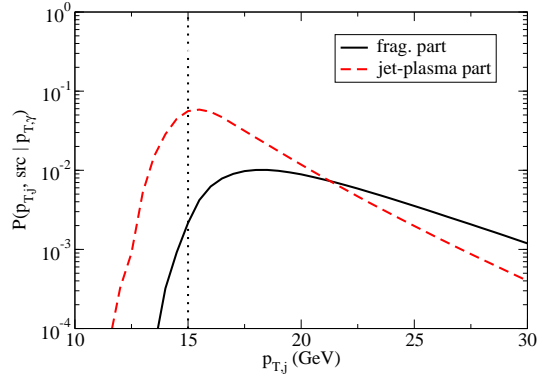} 
\end{center}
\caption{Given a 15 GeV photon as a trigger in central Au-Au collisions, the initial jet distributions are shown. The different contributions from fragmentation, jet-plasma, and direct photons (vertical line) are shown separately. From Ref. \cite{hp2008}.    \label{tagsources}}
\end{figure}
For example, Figure \ref{tagsources} shows the probability distribution of away-side jets tagged by a 15 GeV photon on the near side, before the jet makes it out of the medium and fragments. Given a photon tag, this plot shows the jet probability distributions, for the cases where the jets are tagged by a fragmentation photon, by a photon which originated from an interaction of the jet with the hot plasma, and by a direct photon. This figure shows convincingly that the initial (i.e. before escape and  fragmentation) jet  has more energy on the average than the photon. Also, jets tagged by fragmentation photons dominate the high momentum part of the spectrum, whereas the jet-plasma photons are what tag the lower energy jets except immediately around the tag energy, where direct photons are dominant. Clearly, this information will eventually be encoded in the experimental observables, but the stage shown here is only accessible to the theory. Going a step further from Figure \ref{tagsources}, one may thus ask how this information will influence the measured hadron characteristics: This is shown in Figure \ref{taghadrons}. 
Given a 15 GeV near-side photon trigger in central Au-Au collisions  at the  RHIC top energy, the low $p_{T}$ yield per trigger for hadrons on the away side is dominated by jets whose tags are direct photons. In the higher $p_{T}$ domain, the fraction of hadrons coming from jets tagged by jet-plasma and fragmentation photons is large. Close to the trigger photon energy and higher, the fragmentation photons essentially represent all of the tags.

\begin{figure}[htb]
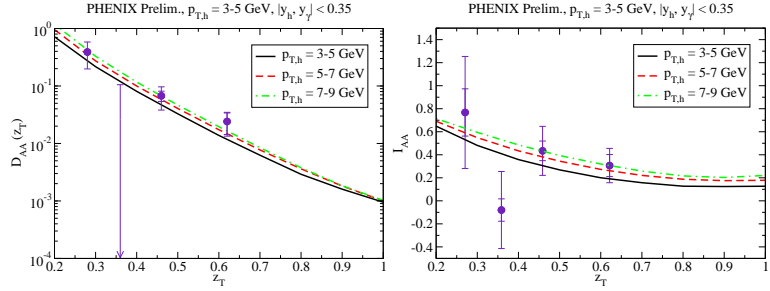

\begin{center}
\includegraphics[width=5.0cm]{phenix-aa-vs-zt-new2.eps} 
\includegraphics[width=5.0cm]{phenix-iaa-vs-zt-new2.eps} 
\end{center}
\caption{(Left panel) The photon-triggered fragmentation function is shown as a function of $z_{T}$, for mid-central Au-Au collisions, at RHIC. (Right panel) The photon-triggered $I_{AA}$ as a function of the momentum fraction, for the same class of events. The data is from \cite{PHENIX-pho}, the calculation from \cite{Qin2009}. Several bins of hadron momenta are shown for comparison.  \label{DAAIAA}}
\end{figure}
\begin{figure}[htb]
\begin{center}
\includegraphics[width=7.0cm]{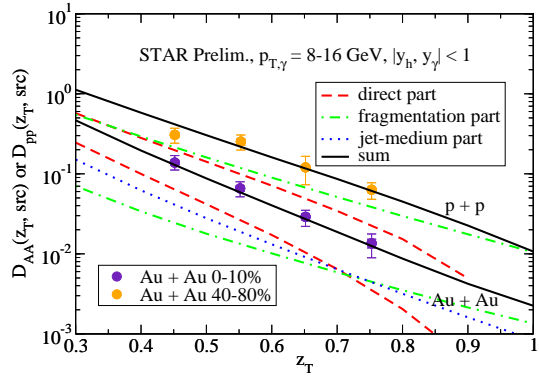} 
\end{center}
\caption{The photon-triggered fragmentation function is shown as a function of $z_{T}$, for central and peripheral  Au-Au collisions, at RHIC. The data is from \cite{STAR-pho}, the calculation from \cite{Qin2009}.  \label{STARDAA}}
\end{figure}

It is exciting that data on photon-tagged hadrons are starting to appear, as the RHIC program presses forward. 
Figure \ref{DAAIAA} (left panel) shows the photon-triggered fragmentation function , $D_{AA} (z_{T})$, as a function of the momentum fraction in mid-central Au-Au collisions at RHIC together with the corresponding experimental measurements from PHENIX. The right panel of the same figure shows the ratio of that quantity to its pp equivalent. This ratio defines $I_{AA} (z_{T})$. Importantly, our approach predicts a flattening of $I_{AA}$ for higher values of $z_{T}$. Note that an accurate treatment of $D_{pp}$ is a  necessary prelude to a successful interpretation of $I_{AA}$. This is the case here \cite{Qin2009}. The STAR Collaboration has also reported photon-triggered fragmentation function m easurements in central and peripheral Au-Au collisions \cite{STAR-pho}. These measurements, together with the appropriate calculations, appear on Figure \ref{STARDAA}.
We find that the STAR measurements span the transition between the jet tagged by the direct photons at low $z_{T}$, to those associated with jet-medium and fragmentation photons. Also, the STAR peripheral measurements are consistent with pp calculations. Clearly,  correlation measurements such as the ones shown in this section have immense potential. On the theory side, a full NLO treatment will put the calculation on an even firmer footing, but results obtained to this point have been more than encouraging: Correlation measurements will represent a large part of the future. Beyond this topical introduction, it is useful to point out that the electroweak tagging of jets can be pursued with lepton pairs \cite{dileptag}, and even with $Z^{0}$ bosons \cite{Z0}. Clearly, the full power of the techniques discussed here with those last observables will only be realized at the LHC. 

\section{Conclusions and outlook}
\begin{figure}[htb]
%\vspace*{-4.2cm}
\begin{center}
\includegraphics[angle=-90,width=7.0cm]{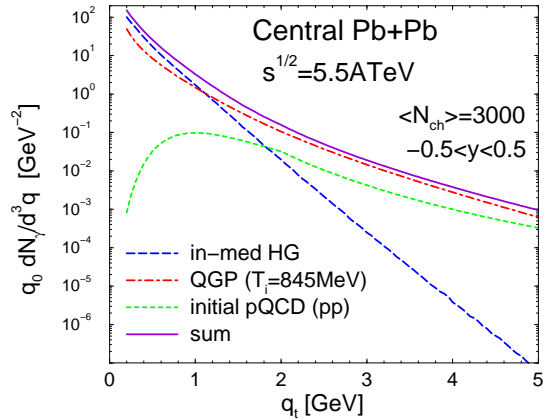}
%\parbox{4.0cm}{\vspace*{4.1cm}\includegraphics[angle=-90,width=5.4cm]{fig2-brem3.ps}}
\end{center}
\caption{Photon spectra predictions at LHC energies. (Left panel) The distribution for soft photons is shown and is together with its different components: That from the quark-gluon plasma (QGP), that from a thermal hadron gas (HG), and that from primordial nucleon-nucleon collisions \cite{TRG}. (Right panel) The panel shows the distribution of harder photons, with its primordial (NN), jet-medium, jet bremsstrahlung, jet fragmentation, and thermal components \cite{Turbide:2005fk}, see also \cite{err}.   \label{LHCsources}}
\end{figure}

It is fair to write that the experimental program on electromagnetic probes in relativistic nuclear collisions has been, and continues to be, a resounding success. Precision photon (and dilepton) spectra have been measured and analyzed at the SPS and at RHIC, and constitute an essential part of the program to be carried out at the LHC. In this context, predictions of photons yields at the LHC have been made for the soft sector \cite{TRG}, and also for harder photons \cite{Turbide:2005fk}. Those two different kinematical regimes are summarized in Fig. \ref{LHCsources} and \ref{BJ_yields}. 
These are but two predictions for LHC energies: A compendium of these can be found in Refs. \cite{LHC,LHC2}. With its impressive increase in available energy, the LHC should venture deep into new territory. It is clear from the figure that the LHC's copious production of jets will feed directly into the photon channels. In fact, even the thermal processes will shine through, at the new collider. Finally, beyond the scope of this text is the story unfolding in hot systems with a high baryon density, such as those currently studied by the HADES dilepton experiment currently at the GSI, and later with the proposed CBM experiment at the same site. 

In what concerns photons produced in relativistic heavy ion collisions, the last decade has seen striking advances in both theory and experiments. We now have photon emissivities complete to order $\alpha_{s}$, for example. Electromagnetic emissivities have also been calculated for a clss of strongly coupled theories, utilizing string theory techniques  \cite{string}. We have also reached a stage where photon measurements can help constrain the dynamical evolution models that are used in the modeling of relativistic heavy ion collisions. One of the reasons for this heightened predictive power is the fact that both ``rates'' and ``evolution models'' have considerably improved in the sense of becoming more quantitative. Together with a sustained effort from the theory community, this progress would not have been possible in the absence of a superb set of experimental measurements of a large palette of complementary variables. Another fact that helps the electromagnetic phenomenology of heavy ion collisions is the fairly recent realization that hard hadronic probes and electromagnetic variables are no longer decoupled. No longer is it possible to treat jet quenching and photon measurements independently. Not only does this coupling provide empirical constraints for the theoretical treatments, but it also serves to enhance further the unity of the field. We have come a long way, and much remains to be done.

\subsection*{Acknowledgments}
I thank all of my coworkers and colleagues for their precious work, insight, and discussions. I am especially grateful to my student and postdoctoral collaborators, and I thank them for their invaluable contribution to much of the material presented here. I thank Reinhard Stock for his support and patience. 
This work was supported in part by the Natural Sciences and Engineering Research Council of Canada, in part by the Fonds Nature et Technologies of the Quebec Government, and in part by McGill University through various student and postdoctoral fellowships.

\end{document}